\documentclass[Afour,sageh,times]{sagej}

\usepackage{moreverb,url}

\usepackage{graphicx}
\usepackage{float}

\newcommand\BibTeX{{\rmfamily B\kern-.05em \textsc{i\kern-.025em b}\kern-.08em
T\kern-.1667em\lower.7ex\hbox{E}\kern-.125emX}}

\begin{document}
\bibliographystyle{plainnat}

\runninghead{Sharan \itshape{et al.} 2024}

\title{Professionalising Community Management Roles in Interdisciplinary Research Projects}

\author{Malvika Sharan\affilnum{1, 2, *},
Emma Karoune\affilnum{1}, 
Vicky Hellon\affilnum{1},
Cassandra Gould Van Praag\affilnum{1},
Gabin Kayumbi\affilnum{1},
Arielle Bennett\affilnum{1},
Alexandra Araujo Alvarez\affilnum{1},
Anne Lee Steele\affilnum{1},
Sophia Batchelor\affilnum{1},
Arron Lacey\affilnum{1,3},
Kirstie Whitaker\affilnum{1, §}}

\affiliation{\affilnum{1}The Alan Turing Institute, London, UK\\
\affilnum{2}OLS (Open Life Science), UK\\
\affilnum{3}Swansea University Medical School, Swansea, UK\\
\affilnum{§}Senior Author
}

\corrauth{\affilnum{*}Malvika Sharan}
\email{msharan@turing.ac.uk}

\begin{abstract}
In this article we discuss community management in interdisciplinary research teams, focusing on recognising and professionalising roles referred to here as the Research Community Managers (RCM). Drawing insights and examples from research and data science projects, we discuss how RCM roles address some of the research's most pressing challenges, from promoting best practices for open research and reproducibility to engaging diverse stakeholders in community-led research and ensuring fair recognition for their contributions. We offer a Community Maturation Indicator and share examples of projects from The Alan Turing Institute, the UK’s national institute for data science and Artificial Intelligence (AI), where institutionally supported RCM roles were established.
\newline
\newline
With the aim to integrate RCM expertise in teams involved in data science and AI research, we provide an RCM Skills and Competencies Framework. We also propose a roadmap for professionalising RCM roles by improving recognition and rewards, potential career paths and organisational support structures. To systematically sustain and progress these roles, we recommend institutional investment in establishing RCM teams that are empowered to prioritise collaboration, transparency and community-based approaches in interdisciplinary projects, such as in data science and AI. As a team, RCMs are well placed to connect disparate teams, initiatives and resources across the organisation, building more resilient research communities that can achieve greater innovation, improved project outcomes and a strongly connected ecosystem, with impacts extending beyond their narrow contexts. 
\end{abstract}


\maketitle

\section*{1 Introduction}
Community management has its roots in the theory of "Communities of Practice" (CoP), a term that first appeared in 1991 to describe the social and informal learning process (also known as ‘situated learning’ (\cite{Lave_1991}) through which novices interact with experts and create professional identities. By the early 2000s, CoPs had transcended their social origins and found application within industries and scientific research communities (\cite{Li_2009}). There are three main characteristics of a CoP: 1) 'domain' that provides shared goals and purpose for members, 2) 'community' that provides a social structure for knowledge exchange, and 3) 'practice', which are resources to address challenges faced by the members (\cite{Li_2009}, \cite{Wenger-Trayner_2015}, \cite{Wenger_2006}). CoPs, spontaneously formed or deliberately designed, foster critical environments for knowledge exchange, deeper engagement and meaningful collaborations that build trust and rapport among participants (\cite{Fadel_2014}, \cite{Hislop_2018}, \cite{Snyder_2010}).
Community management and CoPs play a crucial role in building skills through peer-based learning and adoption of best practices from initial design and development of ideas to knowledge building and information dissemination (\cite{Kulkarni_2000}, \cite{Lesser_2000}). In the context of research, projects that choose to put development and maintenance of material and social infrastructure (also termed repertoire) at the centre of their research work from the start, build resilient and productive collaborations despite the changing requirements of research (\cite{Leonelli_2015}). While everyone in research benefits from CoPs and uses community approaches for collaborations and knowledge sharing, only a few members take active roles in establishing those practices and processes for their entire team (\cite{Borzillo_2011}, \cite{Oliver_1984}). This discrepancy, especially in academic research, stems from inadequate reward for researchers to engage with their communities and overlooked mechanisms to measure their impact on increasing the quality of research outcomes (\cite{Klandermans_1985}, \cite{Sormani_2022}).
Driven by the growing specialisation and diversity of skills required within research and data science, dedicated CoPs, along with community managers to support those communities have emerged in recent years (\cite{SilvaRobles_2017}, \cite{TuringWay_2024b}, \cite{Young_2013}). We refer to these roles as Research Community Managers (RCMs). 

\subsection*{1.1 Definition of Research Community Managers}

\emph{RCMs are professionals responsible for fostering a collaborative environment where a diverse research community can access the socio-technical infrastructure and participatory processes they need to actively engage, get recognition and build shared agency over their work.}

This definition highlights the three core functions of RCM roles:
\begin{enumerate}
\item[(i)] \textbf{Access to socio-technical infrastructure}: RCMs set up the technical infrastructure, ideally with the involvement of the community members, providing access to collaborative tools (such as GitLab and Jupyter) and platforms (such as Discord and Slack). These resources are managed with the goal of improving social interaction, participatory project development, shared documentation and collaboration throughout the project lifecycle  (\cite{Botto_2008}, \cite{Yildiz_2022}).
\item[(ii] \textbf{Facilitation of participatory processes}: RCMs design and organise collaborative activities, community events, workshops, and training sessions to promote active participation from community members. These processes empower users to become engaged members and contributors to different parts of research that they can benefit from (\cite{Racherla_2012}, \cite{Vaughn_2020}).
\item[(iii] \textbf{Building shared ownership and agency}: By ensuring access to skills, resources, and platforms, RCMs allow community members to gain autonomy, build a shared sense of agency and ownership, and contribute meaningfully to the project. They strengthen this by recognising individual and collective contributions fairly and transparently, enabling progress towards both projects and community members’ personal goals (\cite{Baier_1997}, \cite{Bratman_2014}).
\end{enumerate}

\textbf{A note on the term ‘RCM’}: We note that different organisations use different job titles to refer to RCMs, such as community manager, community coordinator, community engagement officer, community outreach manager, community project manager and community educators, among others. Some roles may not even have “community” in the titles, such as network coordinator, project team coordinator, users and developers advocates (used in tech industries) or, simply, research associate or Postdoctoral researcher (traditional academic roles), with responsibilities similar to an RCM. The term ‘RCM’ in this article has been used to encompass all roles that have community-oriented responsibilities in research projects, including in the contexts of data science and AI.
\newline
\newline
In professionaling RCM roles, we draw lessons and inspiration from the Research Software Engineering (RSE) movement, an international initiative aimed at recognising, professionalising, and supporting the role of software engineering in research (\cite{Hettrick_2018}, \cite{Woolston_2022}). In this effort, we have also adopted the naming convention from RSE, using “Research” in the title to both improve awareness and strengthen a shared understanding of community management in academic, government and industry research activities.
\newline
\newline
\textbf{Community management methods and expertise}: Traditional research networks, open source/science projects, and other community-oriented initiatives present opportunities to gain practical expertise in community management on the job or through voluntary roles. Numerous research articles, learning resources and training programmes also focus on developing a theoretical understanding of community management methods to further improve practices in real-world scenarios. These resources are applicable to RCM roles too along with the research experience needed to engage and support research communities. Since the primary focus of this article is on the professionalisation of RCMs, we don't discuss those aspects of community management skills in detail. Nonetheless, we have cited the relevant publications rigorously throughout the article and, for example, purposes, mention a few community consulting organisations that offer learning resources and paid certification programmes for community managers (\cite{CSCCE_2024}, \cite{FEVERBEE_2024}, \cite{TCR_2024}).

\section{2 Understanding How RCMs Evolve Community Building Approaches with their Community}

Each project goes through different stages of development, with each stage influencing the maturation of a CoP developing within it. An RCM adapts community-building strategies at each maturation level to advance a CoP to the next stage.
\newline
\newline
There are several theoretical frameworks, such as Tuckman’s developmental sequence (\cite{Tuckman_1965}) and Peck’s four stages \cite{Peck_1998}, that describe how a community forms, identifies challenges, creates norms and ultimately organises itself. Similarly, Arnstein’s ladder of citizen participation (\cite{Arnstein_2020}), Nabatachi’s increasing level of shared decision authority (\cite{Nabatchi_2012}) and Mozilla Open Leadership’s Mountain/Matrix of Engagement (\cite{Sansing_2019}) provide different perspectives on participation from the citizen, members of public and open source projects respectively. Similar frameworks exist to describe community participation in specific social contexts within which a community exists. 
Although these theories provide good starting points, they do not clearly capture the lifecycle that a research project and, consequently, a research community goes through (\cite{Rosado_2020}, \cite{Sholler_2019}). An RCM considers both the “levels of participation” and “stages of community building” to determine the maturation status of a CoP and inform their approaches for building and nurturing a community successfully.

\subsection{2.1 Introducing a Community Maturation Indicator for RCMs}
\begin{figure*}
\centering
\includegraphics[width=1\textwidth]{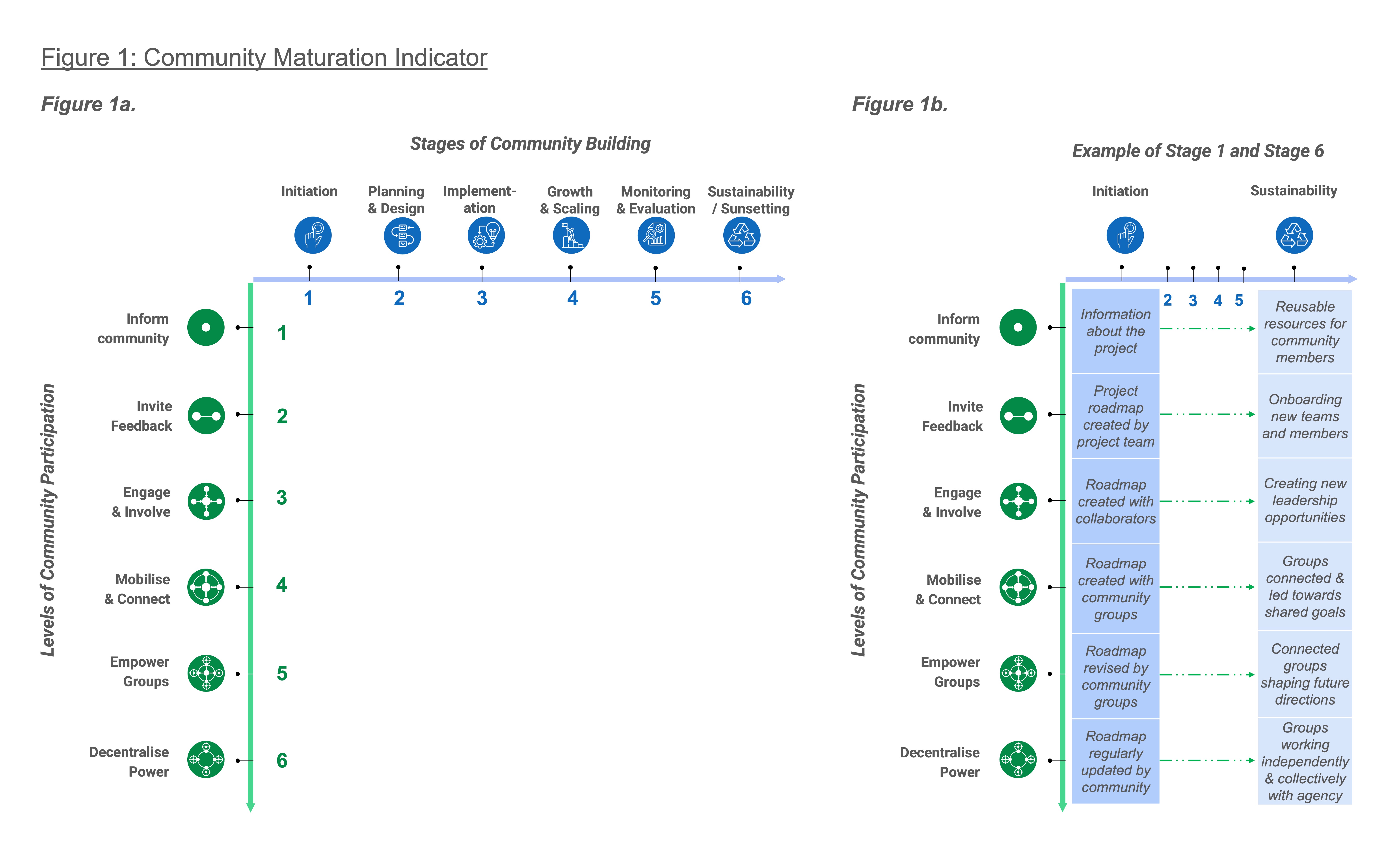}
\caption{
\textbf{Community Maturation Indicator}: 
\emph{This framework for community maturation uses “Level of Community Participation” and “Stage of Community Building” as two main factors to identify the status and inform strategies for community building. 
\textbf{Figure 1a}: The Levels of Community Participation are shown in the x-axis (spanning 6 levels): Level 1: Inform Community, Level 2: Invite Feedback, Level 3: Engage and Involve, Level 4: Mobilise and Connect, Level 5: Empower Groups, and Level 6: Decentralise Power. The Stages of Community Building are shown as y-axis (spanning 6 stages): Stage 1: Initiation, Stage 2: Planning and Design, Stage 3: Implementation, Stage 4: Growth and Scaling, Stage 5: Monitoring and Evaluation, and Stage 6: Sustainability or Sunsetting.
\textbf{Figure 1b}: An example of using a Community Maturation Indicator that shows Stages 1 and Stage 6 of Community Building: “Initiation” and “Sustainability” respectively. For different levels of community participation, from informing the community to decentralising community governance, a community will apply different approaches at all stages. A full table for this example is provided in Supplementary 1.
}}
\label{fig:figure1}
\end{figure*}

Drawing insights from existing frameworks, we have developed a Community Maturation Indicator for research communities in Figure~\ref{fig:figure1}. It illustrates six “Levels of Community Participation” (x-axis/vertical line): 1) inform the community; 2) invite community feedback; 3) engage and involve members in community initiatives; 4) mobilise and connect different groups within the community; 5) empower groups to take different decisions in benefit of the community; and 6) decentralised decision-making power to move away from centrally managed projects.
\newline
\newline
This indicator also illustrates six “Stages of Community Building” (Stages 1 to 6) that RCMs should take into account for CoPs at all levels of participation in their research communities (y-axis/horizontal line): 1) community initiation; 2) planning and design of community strategy; 3) implementation of community strategy; 4) growth and scaling of the community; 5) monitoring and evaluation to build sustainability pathways; and 6) sustaining or sunsetting a CoP.
\newline
\newline
A community can be assigned a specific status at any given time, which can be indicated in the Community Maturation Indicator through a “Level Number - Stage Number” pair (the same stages of development are applied at different levels).

For each “Level of Community Participation”, an RCM has to evolve their approaches to different “Stages of Community Building” before a community can move to the next maturation status (as suggested in the indicator). For different CoPs, an RCM may require different timeframes and resources to move a community from one maturation status to another. Approaches at any maturation status also influence and inform the strategies an RCM can adopt for the next maturation status. A table with descriptions and examples has been shared in \hyperlink{Sup1}{Supplementary 1}. We later use this indicator in Table 1 to communicate the community maturation status for CoPs in different projects.
\newline
\newline
\textbf{Levels of Community Participation}: The level of participation possible within a community depends on several factors, including who the main stakeholders are and what resources are allocated to support them in a CoP. For example, a ‘seminar series' can be informally organised with limited resources to inform interested members on specific topics of interest (Level 1). In action-oriented projects, such as public health projects or policy advocacy, RCMs provide targeted infrastructure and processes through which community members can provide feedback and get involved (\cite{Vasoo_1991}) (Level 2-3). In a formal research partnership such as between academia and industry, an RCM involves specific groups of stakeholders from the partnering organisations, with ideally shared resources provided to engage them (Level 3). To engage stakeholders from different domains, meta-communities (such as in open science) can be established where RCMs mobilise and connect community leaders to collaborate, share practices and combine solutions from their respective domains (\cite{deBayser_2015}). In an open source or open science projects, an RCM engages both project members and volunteer contributors, providing them open and accessible resources that empower them to self-determine their involvement in the community, from using resources to contributing to them or leading new development (Level 4-6) (\cite{Michlmayr_2009}, \cite{Sholler_2019}).
\newline
\newline
\textbf{Stages of Community Building}: After the Level has been identified, a CoP can start applying targeted community-building approaches appropriate for the stage of the CoP. For example, as shown in \hyperlink{Sup1}{Supplementary 1}, at the community initiation stage of a new CoP, an RCM shares information and communicates project goals (Level 1 - Stage 1); at the planning and design stage, an RCM builds stakeholder awareness plans and help to prioritise specific objectives (Level 1 - Stage 2); at the implementation stage, an RCM provides collaborative opportunities for specific stakeholders (Level 1 - Stage 3); at the growth stage, an RCM facilitates knowledge exchange and skill building (Level 1 - Stage 4); and at the evaluation stage, an RCM assesses the success of the CoP through feedback data to inform plans (Level 1 - Stage 5). These stages collectively affect how the project and its community can be sustained in the future or whether it should be paused (sunsetting) (Level 1 - Stage 6). RCMs build effective community strategies and implement creative approaches to support the development of a CoP, advancing a community from one stage to the next, across different levels of participation (\hyperlink{Sup1}{Supplementary 1}, shows possible steps to move from one level to another in the Community Maturation Indicators).

\subsection{2.2 RCMs Adapt Community Building Strategies According to Community Maturation Status}

RCMs build a strong understanding of the technical and cultural contexts within which the communities exist and the shifting power and influence community members can have in the project (\cite{Tritter_2006}). These factors impact the choices and possibilities for infrastructure, platforms and practices adopted by RCMs and their communities. For example, in an open source software project that intends to engage community members in its development, an RCM maintains an online repository under a permissive licence, such as on GitLab or GitHub with MIT Licence (\cite{Saltzer_2020}), where volunteer users and developers can easily become contributors and engaged community members. Whereas, in a public health data project, RCMs may support the use of trusted research environments for data access for authorised members of the project (\cite{Kavianpour_2022}), while also promoting Patient and Public Involvement and Engagement (PPIE) approaches (\cite{Kaisler_2020}). In a participatory project, an RCM may integrate citizen science approaches, inviting members of the public to actively participate as volunteers or paid consultants in research and collaborate with research staff (\cite{Bonney_2014}, \cite{Cohn_2008}, \cite{Kavianpour_2022}). 
\newline
\newline
Similarly, various socio-technical factors influence the evolving needs and demands for community management within a project (\cite{Botto_2008}, \cite{Hislop_2018}). Only through professional recognition, institutional support and appropriate resourcing, RCMs can be empowered to address these changing and often context-based requirements effectively.

\subsection{2.3 RCMs Implement Best Practices to Improve Research Quality}

Quality research refers to rigorous, reproducible, transparent, collaborative, and ethical processes and outcomes, encompassing all aspects of research, from study design to the selection of methods, data collection, analysis, measures against errors or bias (systemic and non-systematic) and dissemination (\cite{Boaz_2003}, \cite{Lohr_2004}, \cite{Margherita_2022}). Responsible Research and Innovation (RRI) underpins research quality achieved through “a collective, inclusive and system-wide approach” through involvement of all stakeholders in the processes of research and innovation, validated through internal and external review (\cite{Hoeven_2013}). This allows research stakeholders to obtain relevant knowledge and resources to inform actions addressing the 'grand challenges' of society and to co-create sustainable research outputs, products, and services (\cite{Hoeven_2013}; \cite{vonSchomberg_2013}, \cite{vonSchomberg_2011}).” 
\newline
\newline
RCMs involvement in research projects can be integral to the operationalisation of RRI’s four dimensions (\cite{NRC_2002}, \cite{Owen_2012}, \cite{Zwart_2014}): i) Anticipation: “[Researchers and innovators] think through various possibilities to be able to design socially robust agendas”. ii) Reflexivity: “[Researchers and innovators] think about [their] own assumptions to consider [their] own roles and responsibilities in research and innovation, and public dialogue”. iii) Inclusion: “[Researchers and innovators] broaden and diversify the sources of expertise and perspectives”. iv) Responsiveness: “[Researchers and innovators] maintain flexibility and capacity to change research and innovation processes according to public values”. 
\newline
\newline
Expertise in research and data science equips RCMs to provide technical support, evidence-led approaches and strategic problem-solving capabilities in research collaboration processes. With multidisciplinary and multi-stakeholder collaborations as a focus of RRI, RCMs further the adoption of best practices and standards that enhance research quality, promoting greater community participation and creating pathways for involving members in addressing research questions (\cite{DiBenedetto_2019}, \cite{Stilgoe_2013}).

Here, we highlight five key areas where RCMs champion best practices and operationalise RRI to improve research quality: 1) Reproducibility, 2) Openness and Transparency, 3) Accountability and Innovation, 4) Equity, Diversity, Inclusion, and Accessibility, and 5) Fair Attribution and Recognition.

\subsubsection{2.3.1 Reproducibility}

For multi-disciplinary collaboration to contribute to scientific progress and research innovation, it is important to integrate tools and methods that lead to reliable and reproducible research (\cite{Claerbout_1992}, \cite{Rethlefsen_2022}). Multiple studies have revealed the "reproducibility and replication crisis" in scientific publications (\cite{Camerer_2018}, \cite{Grahe_2021}, \cite{Ioannidis_2019}, \cite{OSC_2015}, \cite{Wilson_2022}). In the absence of underlying data and methods, a large proportion of scientific experiments and published results can’t be replicated or reproduced by other researchers. In a survey by (\cite{Baker_2016}), more than half of the respondents to the survey (1576 researchers) were unsuccessful at replicating their own work. Researchers must be able to have confidence in their research findings, and those of their collaborators. Failure to apply reproducibility practices, such as by not sharing methods, data and conditions under which results are generated, leads to preventable errors and misleading conclusions for both researchers conducting research, and those attempting to reuse or build on the research outcomes.
\newline
\newline
Designating RCMs to promote practices and skills in reproducibility can build a research culture where community members share responsibilities to create, maintain and sustain reproducible research assets.
\newline
\newline
RCMs build awareness of key reproducibility methods like version control for code and data, data management practices and containerisation for a reproducible research environment, facilitating their adoption in the project. They enhance access to learning resources and training opportunities so that all stakeholders can build skills and apply research reproducibility approaches in their work. Several reproducibility requirements like documentation, collaborative coding, peer review, and code testing inherently serve as excellent approaches for community engagement and involvement \cite{TuringWay_2024b}. These practices also provide onboarding, mentoring and community leadership opportunities among new and established community members. By connecting across different teams within the project, RCMs reduce friction and improve the potential to create a reproducible workflow that incorporates reusable tools, data and algorithms.

\subsubsection{2.3.2 Openness and Transparency}

Open science advances scientific knowledge by promoting transparency across all areas of research (\cite{UNESCO_2021}, \cite{Winker_2023}). By making methods, practices and outputs including all research components open for others, community members can reuse and build upon those components. Open science practices such as open source codebases, open data, open access and open education empower communities to participate in open collaboration, co-creation and sharing of resources \cite{UNESCO_2021}. These are reinforced by reproducibility methods referenced in section 2.3.1, such as version control, data management, licensing, and reproducible workflows, enhancing rigour and reliability in research (\cite{Pownall_2023}). Using the principles of "as open as possible, as closed as necessary," (\cite{EC_2016}) open science can also inform decisions on which aspects of research cannot be made open, with clear justifications, particularly when handling sensitive data or high-risk technology (\cite{Landi_2020}).
\newline
\newline
Advocating for open science can nevertheless be challenging in research teams unfamiliar with these concepts and practices (\cite{Chakravorty_2022}). Researchers may be unsure about securely handling different research objects or determining what components can be openly shared, and to which extent (\cite{Allen_2019}, \cite{Sharan_2020}). RCMs use and share open practices to bridge this gap by guiding open collaboration and open source development approaches across various research facets, all while balancing transparency with awareness of data privacy requirements (\cite{Borghi_2022}, \cite{Morehouse_2024}). They provide accessible project-related resources, community activities, and information in ways that consider diverse accessibility needs and community standards. RCMs lead and educate the community on navigating different scenarios and combine open ways of working. In projects involving sensitive data or which can’t be published openly, RCMs can draw practices from large-scale open source projects but apply them in an ‘inner source’ manner (\cite{Ambler_2000}) to improve collaboration among internal stakeholders (\cite{Izquierdo-Cortazar_2022}). RCMs enhance transparency in the system by acting as bridges between communities, governance bodies and other stakeholders (\cite{Beckham_2023}, \cite{Zaccaro_2002}), sharing information regularly through appropriate channels.

\subsubsection{2.3.3 Accountability and Innovation}

Multi-disciplinary research inherently involves diverse stakeholders with distinct expertise within the project (\cite{Ratcheva_2009}). To build a shared understanding of goals, ways of working and responsibilities, team members must be provided with the resources and information they require to participate. Establishing a project charter with a project vision, mission, roadmap, milestones, stakeholder map and engagement plans early on is crucial in shaping the governance of the project (\cite{Love_2006}). Throughout the project life cycle, it is also critical to ensure interactions and collaboration among all stakeholders promoting comprehensive approaches to problem-solving, use of interdisciplinary methodologies and effective bi-directional flow of information (\cite{Oluikpe_2015}, \cite{Love_2006}). These aspects lead to greater accountability and innovation in multidisciplinary project teams and communities (\cite{Urton_2021}).
\newline
\newline
RCMs enable the creation and centralisation of knowledge from different teams and facilitate communication in an accessible manner for everyone involved. When working with community partners and research stakeholders, they apply participatory approaches, focusing on ethical considerations, clear set of responsibilities and community involvement, all contributing to greater accountability (\cite{FAmauchi_2021}). Through their engagement with different groups, they gain a broad view of the project operations, identifying both roadblocks and opportunities and ensuring timely dissemination of relevant information. Operating at multiple levels, from hands-on tasks to strategic collaboration with senior leadership, RCMs advocate for diverse perspectives and interests from community members, especially in matters that impact them or when they are not directly involved in the decision-making process. By supporting the development of an effective governance process, RCMs can contribute significantly to building accountability among members of a project team and community (\cite{Gardner_2015}). Furthermore, by inviting diverse perspectives via multiple channels for engagement, RCMs support project teams in integrating knowledge from different sources, playing an important role in research innovation (\cite{Khedhaouria_2015}, \cite{Ratcheva_2009}).

\subsubsection{2.3.4 Equity, Diversity, Inclusion, and Accessibility (EDIA)}

Data and AI models, collected, generated, and developed by humans, can be biassed by our social contexts and influenced by factors such as capitalism, ableism, patriarchy, and Western hegemony (\cite{Buolamwini_2023}, \cite{Goffi_2021}, \cite{Held_2023}, \cite{ONeil_2016}, \cite{Verma_2024}, \cite{Zajko_2021}). These biases can cause disproportionate harm to already underprivileged groups and exacerbate societal inequalities (\cite{Birhane_2022}, \cite{Zajko_2022}). One significant source of perpetuating bias is the lack of diversity among researchers, developers, and decision-makers in technology development. Another challenge lies in training datasets that reflect biases in how data were collected and the implicit biases of data collectors, leading to the underrepresentation of diverse needs and perspectives in technologies (\cite{Broussard_2024}, \cite{Verma_2024}). Data science practices that disregard diversity and accessibility considerations create additional barriers for disabled individuals. This limits opportunities to develop technology, for instance, powerful AI models, that are ethically robust, inclusive and equally beneficial for a diverse audience (\cite{Varsha_2023}, \cite{Charitsis_2023}). Responsible and ethical development, deployment, and monitoring of data science and AI systems necessitate the careful implementation of EDIA principles. This involves actively engaging diverse voices in the design, development, and dissemination of rigorous research methods and their outcomes (\cite{Charitsis_2023}, \cite{Leavy_2018}, \cite{Mensah_2023}).
\newline
\newline
Professionals working closer to the community play a crucial role in creating pathways and opportunities for them to contribute their expertise and perspectives (\cite{Arpino_2022}, \cite{King_2024}, \cite{Pompper_2024}, \cite{RCarter-Sowell_2023}). RCMs in research communities operationalise EDIA principles by establishing and enforcing community policies such as community participation guidelines and actionable codes of conduct \cite{Shibuya_2009}, \cite{Sumo_2023}. Through thoughtfully designed stakeholder engagement processes, RCMs create an inclusive environment where individuals from diverse backgrounds and identities, including those likely to be directly impacted by technology, are intentionally included and empowered to influence research directions ()\cite{Bacon_2009}, \cite{Clauss_2018}). RCMs collaborate with a range of experts, including AI ethicists, EDIA specialists, and representatives from underrepresented groups, to develop learning opportunities and resources that integrate EDIA consideration and real-world requirements into research outputs and technologies by design (\cite{Bammer_2020}, \cite{Clauss_2018}, \cite{Medeiros_2023}).

\subsubsection{2.3.5 Fair Attribution and Recognition}

As research in data science and AI becomes more complex and increasingly depends on community contributions, establishing authorship and credits for contributors becomes even more important (\cite{Birnholtz_2006}, \cite{ICMJE_2024}). Assigning appropriate attribution can be challenging, particularly when multiple contributors, each with different roles and levels of interests, are involved (\cite{Haeussler_2013}). Traditional authorship models often fail to capture the diverse contributions of data scientists, software engineers, community organisers, data wranglers, project managers and other professionals (\cite{Rennie_1997}). This becomes even more complicated for projects that involve a broader community of volunteers and external contributors, such as in citizen science (\cite{Vasilevsky_2021}) and open source projects. This can lead to the undervaluation of some roles over others.
\newline
\newline
RCMs are instrumental in developing community processes, guidelines, and policies for community involvement, authorship in traditional research outputs and contributorship across all aspects of a project (\cite{Ozerturk_2021}). Fair attribution and recognition of community-led efforts beyond creating traditional research outputs is essentially an extension of RCM roles in operationalising EDIA principles in community management (section 2.3.4). Their work with community members, while considering individuals' backgrounds, identities, or disabilities, can lead to fair recognition and the creation of opportunities for participation, engagement, and contributions (\cite{Draper_2023}, \cite{Ozerturk_2021}). By highlighting diverse pathways for contributing to ongoing projects, RCMs ensure meaningful recognition and visibility for all contributors (\cite{Birnholtz_2006}). This approach paves the way for future leadership roles within the community, empowering diverse voices and perspectives in shaping project outcomes.

\section{3 RCM Roles in Practice}

\subsection{3.1 Examples of RCM Roles Projects at The Alan Turing Institute}

The Alan Turing Institute (‘the Turing’), founded in 2015, is the UK’s national institute for data science and Artificial Intelligence (AI) (\cite{Turing_2024}. The first RCM at the Turing was recruited in 2019 in The Turing Way, a flagship project supported by the institute (\cite{TuringWay_2024b}). The Turing Way is an open source, open-collaboration, and community-driven initiative for discussing and sharing data science best practices. The RCM role, based on the community manager roles from open source projects, was established to coordinate a distributed team of data science practitioners and the broader network of open science communities. Following its success, several RCM positions across different projects at the Turing have been designed with two different job titles reflecting different experience levels: Research Community Managers and Senior Research Community Managers (\cite{Sharan_2023}). In this article, we use "RCMs" to collectively refer to all community management titles, including RCMs and Senior RCMs at the Turing, unless specifically stated otherwise.
\newline
\newline
The Senior RCMs are experienced community managers who bring some level of leadership experience from research or previous (formal or informal) community management roles. Individuals who may not have prior experience in professional community management roles can be hired as RCMs based on their domain expertise relevant to specific projects and their demonstrated interests and skills in community management. This is often evidenced by their experience in coordinating informal communities of practice or interest groups (illustrated in Figure~\ref{fig:figure5}). 
\newline
\newline
RCMs at the Turing build and nurture CoPs with a diverse set of stakeholders working across different projects in multidisciplinary teams. Since 2021, 12 full-time RCM positions at the Turing have been created to integrate community considerations and best practices into data science and AI projects aligned with the institute’s research priorities and strategic partnerships. These current and previous roles have brought domain-specific expertise relevant to the projects alongside community management experience from their previous roles. Each position is funded through budgets allocated for projects where needs or opportunities for RCMs have been recognised and their roles scoped with the RCM team leads alongside other domain experts. As of August 2024, RCMs have worked across priority areas in health, environment and sustainability, data-centric engineering and skill-building domains, which have been highlighted in Table~\ref{Table1}. We also provide links to project details and case studies that demonstrate different flavours of community management and types of CoPs supported by the RCMs at the Turing.

\begin{table*}
\small\sf\centering
\caption{\textbf{Overview of Projects at the Turing with RCM Team Involvement\label{Table1}}}
\setlength{\tabcolsep}{3pt} 
\renewcommand{\arraystretch}{1.5} 
\begin{tabular}{p{0.02\linewidth} | p{0.12\linewidth} | p{0.27\linewidth} | p{0.05\linewidth} | p{0.07\linewidth} | p{0.15\linewidth} | p{0.07\linewidth} | p{0.07\linewidth}}
\toprule
\#&Project name&Short description&Project Start Year&RCM start year and current status&Community Status on the Community Maturation Indicators&Relevant links&Manuscript authors working on this project\\ \hline
\midrule
    \texttt{1} & \emph{The Turing Way} & Open source        community-led project on data science best practices, featuring an online book organised into six guides & 2019 & 2019 - Ongoing & Empower Groups \textbf{(Level 5)} - Monitoring \& Evaluation \textbf{(Stage 5)} & \href{https://www.turing.ac.uk/research/research-projects/turing-way}{\emph{Details}}, \href{https://github.com/alan-turing-institute/open-research-community-management/blob/main/team-related-comms/briefing-notes/2022-ASG-Briefing_0_TuringWay.pdf}{\emph{Briefing Note}}, \href{https://github.com/the-turing-way/}{\emph{Repository}} & ALS, KW, MS, AB\\
    \texttt{2} & The Turing-Roche Strategic Partnership & Establish a research collaboration between Roche and the Turing to explore patient and disease heterogeneity using advanced analytics & 2021 & 2021 - Ongoing & Engage and Involve Members \textbf{(Level 3)} - Growth and Scaling \textbf{(Stage 4)} & \href{https://www.turing.ac.uk/research/research-projects/alan-turing-institute-roche-strategic-partnership}{\emph{Details}}, \href{https://github.com/alan-turing-institute/open-research-community-management/blob/main/team-related-comms/briefing-notes/2023-RCM-Briefing_Turing-Roche_community-VH.pdf}{\emph{Briefing Note}}, \href{https://github.com/turing-roche}{\emph{Repository}} & VH\\
    \texttt{3} & Turing-RSS Health Data Lab & Independent source of statistical modelling and machine learning expertise to address policy-relevant research questions & 2021 & 2021 - Concluded in 2023 & Engage and Involve Members \textbf{(Level 3)} - Implementation \textbf{(Stage 3)} & \href{https://www.turing.ac.uk/research/research-projects/turing-rss-health-data-lab}{\emph{Details}}, \href{https://github.com/alan-turing-institute/open-research-community-management/blob/main/team-related-comms/briefing-notes/2023-RCM-Briefing_Turing-RSSHealthDataLab_community-EK.pdf}{\emph{Briefing Note}} & EK \\
    \texttt{4} & AI For Multiple Long-term Conditions -Research Support Facility (AIM-RSF) & Connect researchers across the AIM consortia, to ensure the investment delivers long-term, real-world impact & 2021 & 2021 - Ongoing & Invite Community Feedback \textbf{(Level 2)} - Implementation \textbf{(Stage 3)} & \href{https://www.turing.ac.uk/research/research-projects/ai-multiple-long-term-conditions-research-support-facility}{\emph{Details}}, \href{https://github.com/alan-turing-institute/open-research-community-management/blob/main/team-related-comms/briefing-notes/2023-RCM-Briefing_AIM-RSF-OpenCollab_community-EZ.pdf}{\emph{Briefing Note}}, \href{https://github.com/aim-rsf}{\emph{Repository}} & SB, EK, KW, Previously: Eirini Zormpa (EZ) \\
    \texttt{5} & Professional- ising Data Science Roles (Turing's Skills Policy Award) & Collaborate with diverse stakeholders to communicate about and advance the professionalisation of data science roles and skills & 2023 & 2023 - Concluded in 2024 & Invite Community Feedback \textbf{(Level 2)} - Implementation \textbf{(Stage 3)} & \href{https://github.com/alan-turing-institute/professionalising-data-science-roles}{\emph{Details}}, \href{https://github.com/alan-turing-institute/professionalising-data-science-roles}{\emph{Repository}} & EK, MS \\
    \texttt{6} & Turing's Partnership in Innovate UK BridgeAI Programme & Provide independent scientific advice, online resources and tailored training to promote AI adoption among SMEs in the UK & 2023 & 2024 - Ongoing & Engage and Involve Members \textbf{(Level 3)} - Implementation \textbf{(Stage 3)} & \href{https://www.turing.ac.uk/partnering-turing/current-partnerships-and-collaborations/innovateukbridgeai}{\emph{Details}} & AAA, KW \\
    \texttt{7} & \emph{The Turing Way} Practitioners Hub & Engage with experts from partner organisations, supporting the adoption of AI and best practices, including open source and open data approaches & 2023 & 2023 - Ongoing & Invite Community Feedback \textbf{(Level 2)} - Growth and Scaling \textbf{(Stage 4)} & \href{https://www.turing.ac.uk/turing-way-practitioners-hub}{\emph{About}}, \href{https://zenodo.org/communities/the-turing-way-practitioners}{\emph{Case studies}} & MS, AB, Previously: AAA \\
    \texttt{8} & Data Centric Engineering & Strategic partnership between the Lloyd's Register Foundation and the Turing to bring academic and industrial partners together to address new challenges in data-centric engineering & 2018 & 2023 - Ongoing & Inform Community/ Invite Community Feedback \textbf{(Level 2)} - Planning and Design/ Implementation \textbf{(Stage 3)} & \href{https://www.turing.ac.uk/research/research-programmes/data-centric-engineering}{\emph{Details}} & GK \\
    \texttt{9} & Environment and Sustainability (E\&S) Grand Challenge & Use data science and AI to catalyse the next big changes in addressing the climate and biodiversity crisis & 2023 & 2024 - Ongoing & Inform Community/Invite Community Feedback \textbf{(Level 2)} - Planning and Design \textbf{(Stage 2)} & \href{https://www.turing.ac.uk/research/interest-groups/environment-and-sustainability}{\emph{Details}} & CGVP \\
    \texttt{10} & People in Data & Convening a data professionals community to professionalise under-recognised data roles and provide support and training across the UK & 2024 & 2024 - Ongoing & Inform Community/ Invite feedback \textbf{(Level 2)} - Initiation \textbf{(Stage 1)} & \href{https://www.turing.ac.uk/research/research-projects/people-data}{\emph{Details}} & EK \\
\bottomrule
\end{tabular}\\[10pt]
    \begin{center}
          \emph{\textbf{Table 1}: This table summarises various projects at the Turing in which RCM team members are involved. Column 3 indicates the project start year, while Column 4 denotes the year an RCM or Senior RCM was assigned to the project. Column 6 reflects the community maturation status (Level of Community Participation - Stage of Community Building pair) as illustrated in Figure~\ref{fig:figure1}. For each project, the table also provides links to relevant resources such as project webpages, briefing notes, or case studies. The last column provides information about manuscript authors working on this project including previous RCMs involved in the respective projects. A detailed table is provided in the Supplementary 2.} 
    \end{center}
\end{table*}
Successes of RCMs in previous projects have led to greater recognition for these roles in multi-stakeholder project teams. In the long-term strategy published in 2023, RCMs have been attributed by the institute as an important ‘core capability’ alongside the Turing’s Research Engineers Group (REG) (\cite{REG_2024}), Research Application Management (RAM) team (\cite{RAM_2024}) and Data Wranglers (\cite{DW_2024}), each bringing specialised skills in the institute’s data science and AI projects (\cite{Turing_2023}). These roles have been described in detail in related publications (\cite{Karoune_2024}, \cite{Sharan_2024}).

\subsection{3.2 The Turing’s RCM Team Applies System-Level Approaches to Community Management}

\begin{figure*}
\centering
\includegraphics[width=1\textwidth]{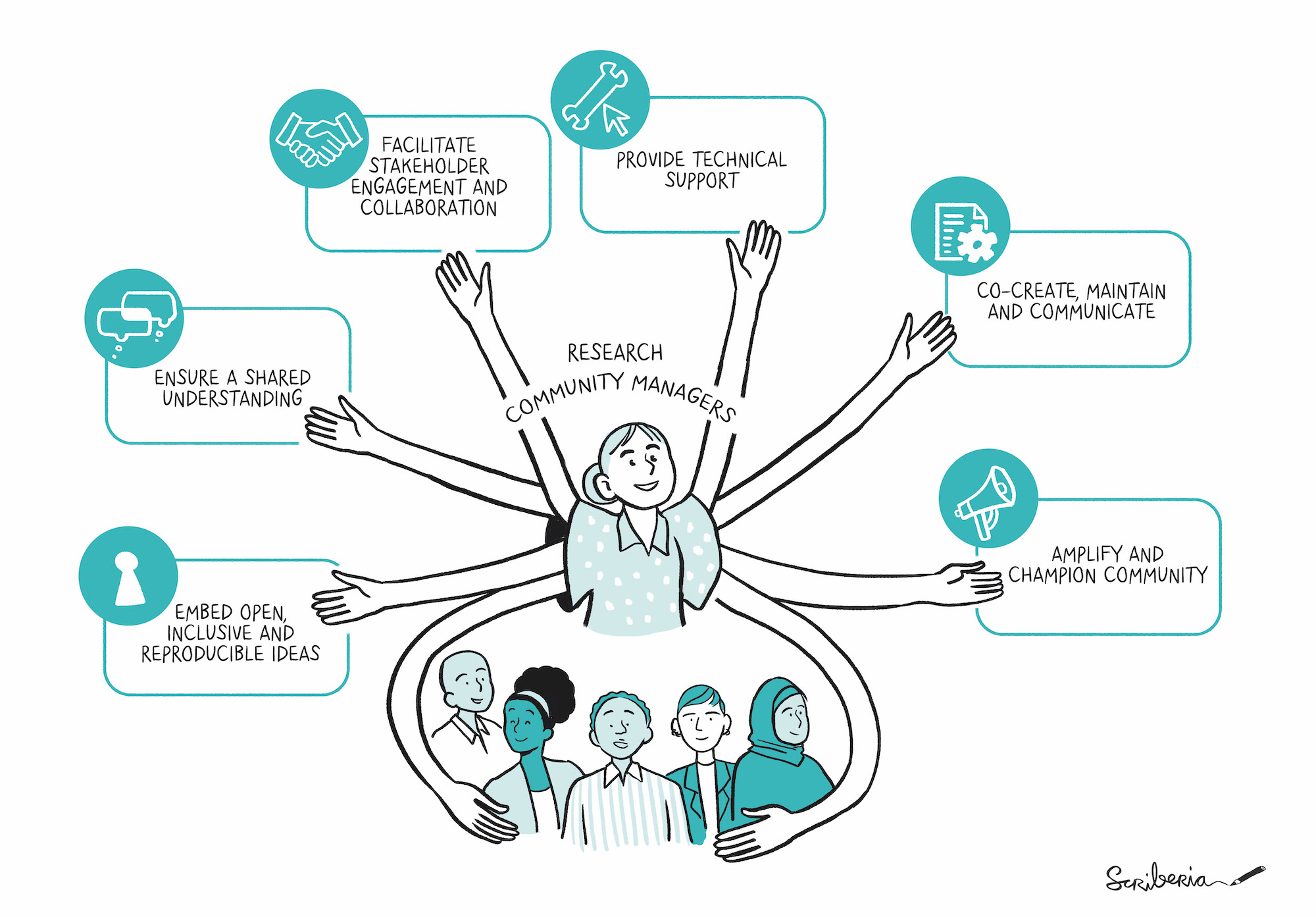}
\caption{\textbf{Overarching Goals of the Turing’s RCM Team}: \emph{The figure outlines the overarching goals of the Turing’s RCM team, including embedding open, inclusive, and reproducible research practices; ensuring a shared understanding of goals, roadmap, and processes; facilitating stakeholder engagement and collaboration; providing technical support and domain expertise; co-creating, maintaining, and communicating project resources; and amplifying and championing community learnings and achievements.}}
\label{fig:figure2}
\end{figure*}

Community management roles are inherently creative, requiring fast-paced, solution-oriented and innovative approaches to navigate emerging challenges and opportunities in communities (\cite{Hughes_2018}). To ensure that RCMs can effectively manage their tasks, they need to be offered appropriate support by institutions (\cite{Barker_2024}).
\newline
\newline
At the Turing, an RCM team provides this support through team-based mentorship, skill building and professional guidance in RCMs’ work (\cite{RCM_2024}). The purpose of membership in a team is to provide opportunities for shared leadership, build standard approaches and establish reusable resources for community management that can be adapted to different projects (\cite{Benishek_2019}, \cite{Martin_2018}).
\newline
\newline
All RCMs take responsibility for CoPs in their respective projects while sharing overarching goals of community management. As illustrated in Figure~\ref{fig:figure2}, the Turing's RCM team have established six goals: 1) embed open, inclusive and reproducible research practices; 2) ensure a shared understanding of goals, roadmap, and processes; 3) facilitate stakeholder engagement and collaboration; 4) provide technical support and domain expertise; 5) co-create, maintain and communicate project resources; and 6) amplify and champion community learnings and achievements.
\newline
\newline
RCMs coordinate their work with each other and apply system-level approaches to connect their projects, project teams, resources and initiatives they are involved in, both at the Turing and from open source/science projects. As close collaborators in the RCM team, RCMs learn from each other’s expertise, refine and amplify their community management approaches and share knowledge from their work regularly, rather than at the end of the project. 

\subsubsection{3.2.1 Turing RCM Team Structure}

Since 2021, the Turing’s RCM team has been led and managed by a Senior Researcher (manuscript author MS), who provides training and mentoring, fosters team knowledge, and offers advisory support across various projects. Over the past three years, the RCM team has been growing in size and maturing its approaches by demonstrating its professional expertise, embedding community-oriented goals across various projects and applying standard community processes and workflows within the Turing. To manage the leadership for the growing team, a Senior RCM was promoted and appointed as the Deputy Team Lead in 2024 (manuscript author EK).
\newline
\newline
The RCM team is operationally housed within the Turing’s Tools, Practices and Systems (TPS) (\cite{TPS_2024}), a research programme, started in 2019 as part of the AI for Science and Government Strategic Priority Fund award (\cite{ASG_2024}). RTP teams, including RCMs and RAMs in TPS, build open source infrastructure to empower a global, decentralised network of people who connect data with domain experts to enable a democratic, accessible and trustworthy research ecosystem (\cite{TuringWay_2024a}, \cite{TPS_2024}). As the Institute’s core capabilities, the RCM team together with RAMs, RSEs and Data Wranglers (\cite{Turing_2023}), embedding open, ethical, collaborative and reproducible data science and AI practices throughout the research and innovation process, both at the Turing and beyond (\cite{DW_2024}, \cite{RAM_2024}, \cite{RCM_2024}, \cite{REG_2024}). The RCM, and other core capability teams, operate in a matrix management style, with line management and professional development sitting in the RCM functional team, while task management day-to-day occurs within research project communities.
\newline
\newline
The RCM team leadership responsibilities include guiding and mentoring the team members, organising team activities and providing community resources. The team leaders support the implementation of proven community approaches and work towards professionalising RCM roles at the Turing and beyond. They are also responsible for developing the team's strategies for short-term and long-term goals and providing a supportive environment where the team members can collaborate to deliver them. The team leaders report to the TPS Programme Director (manuscript author KW). As members of the TPS Programme’s delivery team, they collaborate with other senior researchers and the programme management team (coordinated by manuscript author AB) creating pathways for transparent reporting, advocacy and change at the senior research leadership of the Alan Turing Institute.
\newline
\newline
With the involvement of the team members, the RCM team leaders develop reference materials, connect members of their communities and improve their understanding of RCM expertise more broadly at the institute. They work with members of the RCM team to extend the impact of community management through research and other cross-theme projects, as well as provide consultation to different teams at the Turing. They engage the team in peer-mentoring activities for embedding open science, reproducibility and ethical approaches into their work. They create centralised resources and offer training on best practices, such as open source, reproducibility, collaboration, community engagement, and capacity building in research and data science.
\newline
\newline
To operationalise the Turing's values and deliver on its strategic goals, the team has developed guiding principles that are applied across all their work. They support and guide each other in building the expertise required in their work as well as achieving the team’s goals. They use weekly meetings and co-working sessions as a consistent space to address specific challenges from their projects, exchange resources and discuss innovative approaches for community management. The team members keep each other updated on their work, share emerging ideas and collaborate in developing solutions to address shared challenges. They maintain documentation, practical toolkits, generalised practices, reusable templates and impact reports with examples from their projects, openly shared via the team's GitHub repository (\cite{RCMGitHub_2024}).
\newline
\newline
The RCM team has also been building a professional identity for RCM roles, developing research-based resources and communicating about the RCM roles at the institute and conferences nationally and internationally. The team members actively engage with meta-communities and external open science projects, learning from others’ research processes and sharing outputs from their work openly, such as in The Turing Way. This paper itself is a result of the collaboration and knowledge exchange within the RCM team.

\section{4 Introducing an RCM Skills and Competency Framework to Support the Professionalisation of Research Community Management Roles}

\subsection{4.1 Understanding the Need for Professionalisation of RCM Roles}

Professionalisation is a process of recognising an occupation, which involves specific skill sets, prolonged training processes and established qualifications for improving the effectiveness of a profession. This process leads to raising the status of the profession, giving a professional a certain degree of agency and autonomy over their work, and compensating fairly for their service (\cite{Abbott_1991}, \cite{Horn_2016}, \cite{MariaRosaria_2000}).
\newline
\newline
Although RRI integrates inclusive approaches for community participation and engagement, the responsibility for community management has mostly remained informal, with appropriate recognition and support for RCMs and similar roles often lacking (\cite{Glerup_2014}). These roles are highly important in research (\cite{Eder_2012}, \cite{Lusk_2019}), yet they are frequently undermined and remain ‘hidden’ when communicating about research outcomes and impact. This results in RCMs operating without any formal title, certainty of career paths or adequate resources for their work. Therefore, an important step towards empowering researchers skilled in community management is to professionalise RCM roles and sustainably advance their careers while recognising their skills in the research ecosystem (\cite{Branka_2016}).
\newline
\newline
Combining our insights from open science initiatives and research organisations, we focus on the frameworks needed for the professionalisation of RCM roles within the broader research and data science ecosystem. Transitioning RCM roles from informal to formal occupations will present a structured path where researchers' contributions to community management can be openly acknowledged and effectively utilised. 
\newline
\newline
In this effort, we first present the 'RCM Skills and Competency Framework' as a reference to integrate RCM roles and expertise into research and data science projects. We then describe important steps towards professionalising RCM roles across four categories: 

\begin{itemize}
    \item Articulating skills and competencies
    \item Aligning recognition and reward systems
    \item Building long-term career development pathways
    \item Offering prolonged mentorship and support to enable future growth opportunities
\end{itemize}

In the era of data science, we view the professionalisation of RCMs as a crucial part of an important culture change, enabling diverse research and research infrastructure to be supported, incentivised, and empowered for their significant contributions to team science \cite{Eder_2012}. By positioning RCMs at the interface of various stakeholders, including diverse Research Technical Professionals (RTP) (\cite{UKRIEPSRC_2024, UKRISTP_2024}), this shift will prioritise community goals, maximise outcomes and enhance the real-world impact of interdisciplinary research projects.
\newline
\newline
Resources provided in this article can be used as a structured roadmap for fostering the growth and integration of community management in research and data science organisations with appropriate standards for training and professional accreditation of RCM roles.

\subsection{4.2 RCM Skills and Competency Framework}

\begin{figure*}
\centering
\includegraphics[width=1\textwidth]{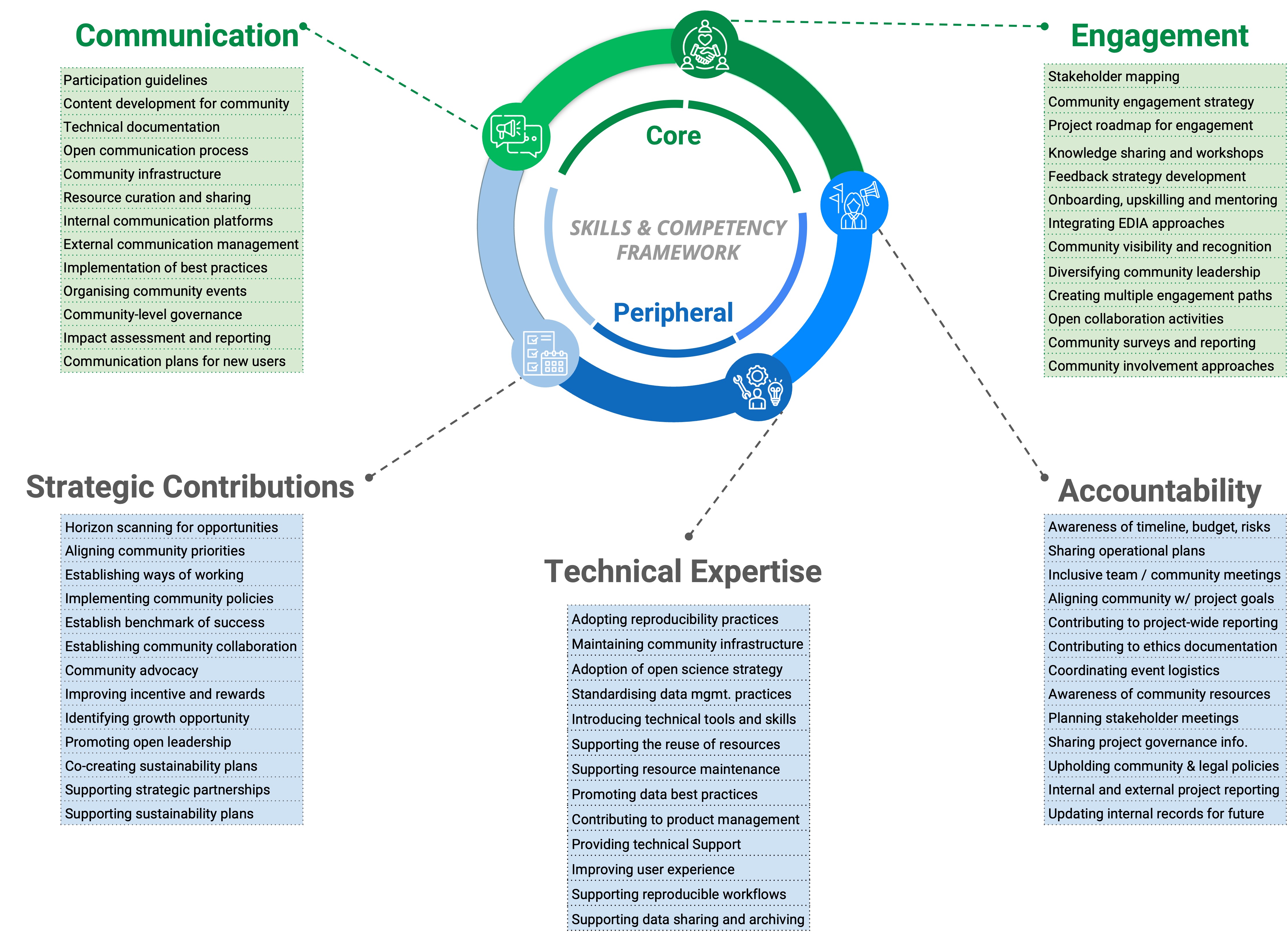}
\caption{\textbf{RCM Skills and Competency Framework}: \emph{This figure illustrates a framework comprising five overarching competencies, described with 13 skills each. A total of 65 skills are categorised into core competencies in communications and engagement, and peripheral competencies in strategic contributions, technical skills, and accountability. A table with details has been provided in Supplementary 3.1}.}
\label{fig:figure3}
\end{figure*}

The RCM Skills and Competency Framework in this article discusses five overarching skills areas needed to perform the RCM roles: i) communications, ii) engagement, iii) strategic contributions, iv) technical skills, and v) accountability (Figure~\ref{fig:figure3}; \hyperlink{Sup3-1a}{Supplementary 3.1}).
\newline
\newline
As illustrated in Figure~\ref{fig:figure3}, a total of 65 skills have been outlined to assess the application of these five skill areas, categorised under core and peripheral competencies. Communication and engagement skills are essential core competencies for RCM roles, with responsibilities carried out by RCMs at a high level of proficiency and expertise. The remaining three skill areas can be considered peripheral competencies—RCMs bring knowledge and awareness in these areas, with responsibilities often carried out in collaboration with other team members, including project leaders, domain experts, and other RTPs or specialists, to meet specific project requirements.
\newline
\newline
We draw from our experience working with community initiatives, community-oriented research projects, informal networks and formal partnerships in data science projects. As researchers and community members in different disciplines, the 10 current and past members of the Turing's RCM team have collectively participated in over 50 different projects involving research communities at the Turing, other research organisations and the broader open science communities. Two lead authors have also established community-based non-profit organisations and international projects promoting open science practices in research. In conceptualising the framework, we combine our experience from working with these communities and insights from existing research publications and resources referenced throughout this article. Selected examples from the Turing are provided in Table~\ref{Table1}.
\newline
\newline
Community Roundtable and CSCCE have previously published skills frameworks for community managers in industry and scientific research respectively (summarised in \hyperlink{Sup3-2}{Supplementary 3.2}). The Community Roundtable's Community Skills Framework, developed in 2014 using survey data from online technical communities, identified 50 skills grouped into five families: content, technical, business, engagement, and strategic (\cite{TCR_2014}). CSCCE's 2021 framework, based on research community surveys, outlines 45 skills categorised into five core competencies: communication, technical, interpersonal, program management, and development (\cite{Woodley_2021}). The RCM Skills and Competency framework presented in this article has been aligned with these resources, while primarily focusing on the skills and competencies that RCMs apply in interdisciplinary research and data science projects.
\newline
\newline
The RCM Skills and Competency Framework include community management skills and responsibilities undertaken in both formal and informal capacities. This includes community roles fulfilled by paid staff, appointed community representatives and members from a volunteer community, whose responsibilities may overlap with other specialised roles within interdisciplinary teams. This framework adheres to the principles of "as open as possible, as closed as necessary (\cite{EC_2016}),” while intending to promote responsible open research (open source, open data and other areas of open science) and reproducibility practices in data science projects.

\subsubsection{4.2.1 How to use this framework?}

The RCM Skills and Competency Framework serves as a tool and reference for various audiences within data science and AI communities, including but not limited to the following:
\begin{itemize}
    \item \textbf{Organisations}: Establishing the infrastructure needed to develop and sustain RCM roles and RCM teams.
    \item \textbf{Experienced community professionals}: Enrich their own skills or support other community managers in identifying skills and competencies they want to develop.
    \item \textbf{New community managers}: Understanding their own fit as an RCM and career development opportunities in community management.
    \item \textbf{Individual researchers}: Gaining insight into what RCM roles are and how they complement other specialist positions within a research team.
    \item \textbf{Research teams}: Considering the integration of community approaches and delegating responsibilities among team members with different areas of expertise in the projects.
\end{itemize}

A shared understanding of these skills and competencies can guide the creation of new research positions and job descriptions, prioritising areas most crucial for fulfilling the project's goals. 
Specifically, the classification of core and peripheral competencies will help identify skills and responsibilities for RCMs that they carry out either independently or in collaboration with other members of a multi-stakeholder or interdisciplinary project team.

\subsection{4.2.2 Core Competencies: Communication and Engagement Skills}

Collaboration, through communication and engagement, lies at the heart of the RCM role. In our work, we identify communication and community engagement as core competencies for RCM roles. With a high level of expertise, RCMs are responsible and accountable for applying different skills across these two competencies in order to drive successful collaboration and outcomes of the project. From facilitating community inception to nurturing its growth and sustainability, RCMs leverage their specific skills in these competencies to create a transparent, two-way collaboration system with community members. The communications and engagement skills enable RCMs to fairly represent and advocate for the community's interests throughout the research project, ultimately leading to enhanced community awareness, collaboration and involvement in project outcomes.

\subsubsection{\textbf{4.2.2.1 Communications Skills}}

RCMs work to enhance transparency by making implicit knowledge explicit and making information accessible through effective communication. Documentation serves as a key communication resource for the community, ensuring clear visibility of community norms, processes, and outcomes. This includes communicating the vision, roadmap, milestones, and accomplishments of the community. While documentation is a common reference point, synchronous communications allow RCMs to facilitate more informal and dynamic engagement. RCMs set up community platforms to connect and allow cross-pollination between members and stakeholders, enabling them to navigate various aspects of the community, including engagement pathways, contribution types, modes of recognition and sustained collaborations. Beyond communication and knowledge sharing, RCMs actively shape the community experience by designing community programs and events, overseeing community-level governance processes and facilitating inreach and outreach initiatives. RCMs align the organisational impact assessment process with a community’s communication strategy to effectively report on community success. They collaborate with the community to improve the accessibility of resources through translation, contextualisation, and localisation. Additionally, RCMs also create communication materials to promote the community's work to non-specialist audiences or funding partners.

\paragraph{\textbf{Examples from the Turing}}
All RCMs at the Turing develop resources to communicate about their work and invite feedback by sharing them with their respective communities. For example, in The Turing Way (Table~\ref{Table1}, project 1), the RCM (manuscript author ALS) drafts monthly newsletters with updates from various communication channels and sources, disseminating them to all community members, including those not directly involved in the project. Other communication materials, such as GitHub documentation, presentations shared on Zenodo and Slack or social media announcements enable all members to stay informed or locate specific information they need. The Turing Way handbook includes a Community Handbook that shares community-related guidance and resources for anyone to read and reuse in their work (\cite{TuringWay_2024b}). The RCM also collaborates closely with different members and working groups in the community, ensuring effective communication through standard channels and contributing to the decision-making process. For example, the RCM leads the Collaboration Café, a biweekly online collaborative meeting, where members working on various parts of the project can come together and invite feedback through active collaboration during and beyond these calls. The RCM also supports community members in communicating about The Turing Way through international conference talks, training workshops, and the project’s twice-yearly Book Dash events, which facilitate synchronous work across different chapters of the book. 
\newline
\newline
All RCMs at the Turing aim to support a bidirectional flow of knowledge by curating and adopting data science best practices from The Turing Way in their work and sharing practices and examples from their work in The Turing Way handbook. Through their efforts in the RCM team, they develop reusable resources and communication materials and cross-post them in The Turing Way.

\subsubsection{\textbf{4.2.2.2 Engagement Skills}}

Research projects thrive not just on scientific questions, but on their social and technical infrastructure. This socio-technical infrastructure provides a space for researchers to connect around shared goals and visions. RCMs apply community engagement skills with socio-technical considerations to map stakeholders, providing diverse engagement pathways and maximising the positive impact of research. They ensure inclusivity and diversity by aligning community and organisational policies with EDIA principles. RCMs establish clear processes and roadmaps for community onboarding and engagement, fostering open and inclusive spaces like events, documentation, platforms, and forums. These spaces act as catalysts for collaboration, skill exchange, and project improvement. Essential skills for RCMs include understanding community needs and personas, designing onboarding processes for long-term collaboration, and ensuring project sustainability. They collect data, perform community assessments, and measure, and report on community health, using these insights to continuously improve processes and infrastructure.

\paragraph{\textbf{Examples from the Turing}}

Ambassador programs are effective at engaging specific community members through structured, time-bound, and value-based activities. An example at the Turing is the Turing-Roche Scholar Scheme (\cite{Turing_Hellon_2023}), developed and run by the RCM for the Turing-Roche Strategic Partnership (Table~\ref{Table1}, project 2, manuscript author VH). The scheme supports 10 UK-based PhD students to embed themselves with the Turing-Roche partnership, a unique academic-industry collaboration, and helps them develop skills that will benefit and further their careers. Scholars are mentored and supported to undertake a community-based project of their interest, representing topics relevant to the partnership and data science and health more broadly and allowing the partnership community further opportunities to engage. Scholars also receive regular group cohort calls throughout the year are given a stipend and attend a relevant conference sponsored by the partnership. This program incorporated advice from other ambassador programs such as the Software Sustainability Institute Fellowship (\cite{SSI_2024}) and the OLS (formerly Open Life Science) cohort-based training and mentoring program (\cite{OLS_2024}).
\newline
\newline
Another example is from Innovate UK's BridgeAI program (Table~\ref{Table1}, project 6), which supports the development of AI skills among Small- and Medium-sized Enterprises (SMEs) in the UK (\cite{Turing-BridgeAI_2024}). The project's Senior RCM (manuscript author AAA) is responsible for engaging different community stakeholders, including Turing teams involved in the project, external advisory members who provide sector-specific consultation for SMEs, and the broader network of BridgeAI partners. Through structured meetings and community calls with members from across these teams, the Senior RCM facilitates discussions, highlights collaboration opportunities, and shares resources from asynchronous collaboration on different project aspects. These interactions have enabled the Senior RCM to inform plans for the next stages of the project, such as the recruitment of future advisory groups, co-development of useful resources for SMEs and community calls. The Senior RCM engages with community members and invites their contributions to co-designing, leading, and participating in stakeholder engagement activities such as workshops, panels, and webinars. These community engagement efforts keep members informed and aligned while avoiding duplication of efforts.

\subsubsection{4.2.3 Peripheral Competencies: Skills for Strategic Contributions, Technical Expertise, and Accountability}

RCMs often start their careers on projects with specific community management needs tied to project goals. However, as they gain experience across diverse projects and communities, their understanding of the broader functioning of the project and data science evolves, influenced by the scope and scale of the communities they lead or contribute to. The breadth of information allows RCMs to foster a systems-level understanding of community roles and their interconnectedness with individuals, projects, and the organisation's ecosystem. To encompass the generalist knowledge of RCMs beyond their specialist roles, we have identified peripheral competencies encompassing three areas: strategic contributions, technical expertise, and project management.
\newline
\newline
While some skills within these competencies are crucial to RCM responsibilities, they often consult or collaborate with other specialists to fulfil the requirements necessary for the project's and community's success. In doing so, RCMs stay informed, offer expertise, and contribute as needed in delivering on specific goals and objectives within a project. Specifically, RCMs work with the leadership team, domain experts, RTPs and other members with formal or informal roles such as RSEs, RAMs, Data Scientists, Research Project Managers (RPM), Data Wranglers, UX specialists and Data Stewards (\cite{Karoune_2024}, \cite{Sharan_2024}, \cite{TuringWay_2024a}). RCMs also consult other institutional teams such as partnerships, outreach and communications teams, or external partners involved in the project team.

\subsubsection{\textbf{4.2.3.1 Strategic contributions}}

RCMs in most projects are required to bring two levels of expertise: 1) to enable operational activities required at the grassroots of the community, and 2) to influence and inform strategic decisions made by the project leadership. This requires balancing community-facing activities with strategic planning and development. Strategic RCM skills include horizon scanning, crafting communication/engagement strategies, and building community roadmaps. They foster open leadership by establishing fair governance (like policies, Code of Conduct, and incentives) and aligning institutional strategy with community goals. They connect with members of their communities and advocate for their interests at the institutional level. Senior RCMs manage staff members or a team, develop funding and sustainability plans, engage in advocacy across all areas of the project internally and externally, and contribute to strategic decisions like project expansion or sunsetting. Operational activities like workshop planning, feedback gathering and transparent reporting fall under communication and engagement skills, but these activities also play a crucial role in strategy by providing insights to measure success, assess community health, communicate impact and identify programming gaps. This dual nature requires RCMs to switch contexts constantly, learning new skills to ensure community and project sustainability. To deliver on this creative yet demanding aspect of their role, RCMs work closely with other RTPs and infrastructure roles such as RSEs, Data Wranglers, and RAMs to support various types of work in research communities and enhance their impact.

\paragraph{\textbf{Examples from the Turing}}

The Turing-RSS Health Data Lab (Table~\ref{Table1}, project 3) was established in 2020 through a partnership between the Turing and the Royal Statistical Society. In collaboration with the UK Health Security Agency (UKHSA), the Data Lab provided independent statistical modelling and machine learning expertise for pandemic response. The Data Lab focused on policy-relevant interdisciplinary research, addressing issues such as social inequalities in COVID-19 risk, debiasing testing data, assessing acoustic markers for diagnosis, and using wastewater as a local prevalence biomarker. The project’s Senior RCM (manuscript author EK) brought substantial research experience and provided community leadership, bringing together researchers from across different teams. Under her coordination, the Data Lab achieved significant milestones in under two years, including completing seven projects, publishing six peer-reviewed articles, and producing three preprints, while embedding open source, open access, open data and reproducibility practices in the Data Lab’s work. In close collaboration with different project members, the Senior RCM also created technical and non-technical project reports, enhancing the accessibility and impact of the research. She designed and hosted knowledge-sharing events and an international public lecture series, strengthening collaboration between senior leaders and early career researchers, and engaged the international public with the project’s focus. The Senior RCM offers specialist consultation for various teams at the Turing, including special interest groups, the Academic Programme, the Skills team, and business teams. This additional responsibility led to her transition into a more strategic community management role within the Turing’s Health Programme, followed by her promotion to Deputy Lead of the RCM team.
\newline
\newline
Another example is from the Turing’s Data-Centric Engineering (DCE) program (Table~\ref{Table1}, project 8), where the Senior RCM (manuscript author GK) works closely with a Research Application Manager (RAM), each with distinctive responsibilities but complementary expertise in the project. One of the projects within DCE is ADViCE, the Artificial Intelligence for Decarbonisation’s Virtual Centre of Excellence. The broader ambition of the Centre of Excellence is to coordinate and engage with AI and decarbonisation stakeholders across high-emitting sectors, and it is delivered by a collaborative consortium including Digital Catapult, Energy Systems Catapult and The Alan Turing Institute. At the strategic level, this initiative aims to foster cross-sector collaboration, define key challenges that can be addressed with AI solutions, and disseminate information to relevant stakeholders. Working closely with the RAM, the Senior RCM has developed the communication infrastructure to support ADViCE’s community engagement and future collaborations. The RCM-RAM pair has implemented the ADViCE Knowledge Base (\cite{ADViCE_2024}) and Forum (GitHub Discussions) which facilitate knowledge sharing within the targeted communities openly and collaboratively. In addition to building the community infrastructure, their deployment into the project includes engaging with ADViCE’s stakeholders by leading knowledge-sharing convening events (for example, data science story-sharing circles). 

\subsubsection{\textbf{4.2.3.2 Technical expertise}}

While RCMs share core community management skills, interdisciplinarity and domain-specific expertise may be crucial for some activities. This often translates to applying technical skills in managing both community responsibilities and meeting relevant technical needs in the project. RCMs bring academic or industry experience within the project's domain, often due to their professional backgrounds in research and data science. Leveraging their combined skillset, RCMs offer technical support and implement research best practices. They build and maintain community infrastructure to streamline access for members to the skills, resources, and infrastructure they need. They also actively encourage and celebrate diverse contributions to address domain-specific and technical challenges. In partnership with other RTPs and infrastructure roles such as RSEs, Data Wranglers and Data Stewards, RCMs establish essential infrastructure such as version control systems, communication platforms, and open science practices encompassing open source, open access and open and FAIR (Findable, Accessible, Interoperable, Reusable) data (\cite{Wilkinson_2016}). By delivering training and workshops on collaborative coding, code review, data analysis, user testing, and data management, RCMs empower the community and contribute to its long-term sustainability.

\paragraph{\textbf{Examples from the Turing}}

RCMs at the Turing have research experience and technical skills that they apply to manage their tasks and workflows. For example, RCMs conduct community research to gather qualitative data through 1:1 interactions and community interviews, and quantitative data to inform their community development strategies. One example is stakeholder data mapping, a critical piece of work through which RCMs identify community stakeholders and their levels of engagement in the project. In smaller projects, this mapping can be achieved easily using simple data tables or stakeholder prioritisation criteria. However, large-scale initiatives like the Environment and Sustainability (E\&S) Grand Challenge at the Institute (Table~\ref{Table1}, project 9), involve mapping stakeholders from environmental research across the UK and internationally. The E\&S Senior RCM (manuscript author CGVP) applies technical skills to map stakeholder data, which is a crucial first step towards building strategic engagement and community-involved plans. She has created a workflow to gather public and individual consented information, harmonise data from different sources and analyse them using network analysis and programming techniques (\cite{GouldvanPraag_2024}). To effectively communicate her findings with stakeholders, she visualises the stakeholder relationships using Kumu, an open source platform that offers an interactive map of complex relationship data (\href{https://cassgvp.kumu.io/alan-turing-institute-environment-and-sustainability}{Kumu demonstration}). She follows the Turing’s data protection plan and stores data under the UK (General Data Protection Regulation) GDPR. Code and guidelines have been shared for reference and reuse on a \href{https://github.com/alan-turing-institute/environment-and-sustainability-gc-community/tree/main/docs/stakeholder-mapping}{GitHub repository}.
\newline
\newline
RCMs also apply their technical skills in supporting their community members and research-related project goals. An example of this can be explored in the AI for Multiple Long-Term Conditions (AIM) programme (Table~\ref{Table1}, project 4) (\cite{AIM-RSF_2024}). Funded in 2021, AIM encompasses seven consortia spanning 28 universities, 12 NHS trusts, and numerous charities, local government bodies, public organisations, and healthcare providers across the UK. The AIM’s Research Support Facility (RSF) facilitates collaborative research within the programme, focusing on five key themes: secure and interoperable infrastructure, research-ready data, open collaboration, public and patient involvement and engagement (PPIE) (\cite{Kaisler_2020}), and sustainability and legacy. Working alongside senior researchers, Data Wranglers and RPMs, two RCMs provide community management under two core themes: Open Collaboration (previous team member Eirini Zormpa) and PPIE (manuscript author SB). Their technical responsibilities include building and maintaining technical infrastructure such as a GitHub organisation, and promoting specialised skills including data standards, public engagement, reproducibility, and open science. The RCM responsible for PPIE developed a 'glossary of terms' to simplify technical concepts, delivered training to upskill patients on research and data science methods and hosted community calls focused on PPIE, making it easier for the public to both understand and integrate their perspectives into the research program. Meanwhile, the RCM for Open Collaboration delivered a series of training workshops introducing programming languages, open source and FAIR practices and reproducibility techniques to enhance the skills of Early Career Researchers. Collaboratively, the RCMs adapted The Turing Way’s Collaboration Café, providing a platform for connection, knowledge exchange, and the adoption of standardised approaches across AIM projects.

\subsubsection{\textbf{4.2.3.3 Accountability}}

RCMs employ strong management and leadership skills to build accountability in the community. They facilitate the access and use of community resources ethically and equitably, guiding the community towards shared goals in the project. These skills overlap with the project leadership and those in dedicated project management roles, like RPM or business administrators. RCMs' skills complement rather than compete with these roles and their responsibilities. The project lead oversees decision-making and ensures that the project progresses according to its goals and requirements. RCMs provide a deep understanding of community needs to inform the project leadership and shape project management decisions. They fulfil this by; regularly sharing updates from the community; supplying community documentation; highlighting areas where additional management input could benefit the community; clarifying project-level governance and its connection to community governance; and prioritising tasks to streamline information flow between the community, research team, and project leadership. RPMs on the other hand remain responsible for keeping track of the project timelines, budgets and operational plans, establishing clear reporting structures, identifying risks and mitigation strategies, and coordinating finances and recruitment. In close collaboration with RPMs, RCMs clarify their focus areas and integrate community-centric approaches into project delivery.

\paragraph{\textbf{Examples from the Turing}}

Many projects at the Turing allocate RPM support with whom RCMs actively collaborate to achieve project goals and ensure that their work follows institutional policy and legal requirements. Given some overlaps that RCMs have with RPMs, project and community responsibilities are openly discussed at different stages of a project. This close collaboration is instrumental in aligning priorities from the project and related community while building accountability. All members of the RCM team who work with RPMs on their projects have reported feeling better supported and experiencing improved well-being at work. Having dedicated RPM support has led to more successful project outcomes, timely project completion, and a more manageable workload by allowing RCMs to focus on their core responsibilities. This close collaboration has also been beneficial for RPMs in their professional growth. 
\newline
\newline
A good example of RPM-RCM collaboration can be illustrated in The Turing Way’s Practitioners Hub (Table~\ref{Table1}, project 7), a project under the BridgeAI program that engages organisations such as Small and Medium-sized Enterprises in adopting AI skills while leveraging open source and open data practices to improve their businesses. The Practitioners Hub offers time-bound (4-6 months) cohort-based activities to work with specific groups of partnering organisations and SMEs, who dedicate part of their work time to engage with the programme. The first cohort, delivered in 2023, included six training workshops, multiple informal meetings, two high-impact cross-sector events and six case studies co-developed with the cohort participants. A Senior member of the RCM team (manuscript author MS) is responsible for the strategic planning and maintaining collaboration with the cohort members and participating organisations, while a high degree of RPM involvement has been critical for the successful delivery of the project. The RPM provides support in managing timelines, budgets, and legal contracting in coordination with the institute’s partnership, legal, ethics, events and members of the project team. Working closely with RCM team members, such as when hosting professional workshops, organising community events and planning cohort engagement activities, the RPM handled the operational requirements. The first RPM (manuscript author AAA) in the project also gained new skills and insights into the RCM’s workflow, which led them to later join the RCM team as a BridgeAI Senior RCM (Table~\ref{Table1}, project 6). In the absence of an RPM, a Programme Manager (manuscript author AB) provides such support, while handling similar responsibilities across multiple projects.

\section{5 Recommended Steps Towards Professionalising RCM Roles}

\begin{figure*}
\centering
\includegraphics[width=1\textwidth]{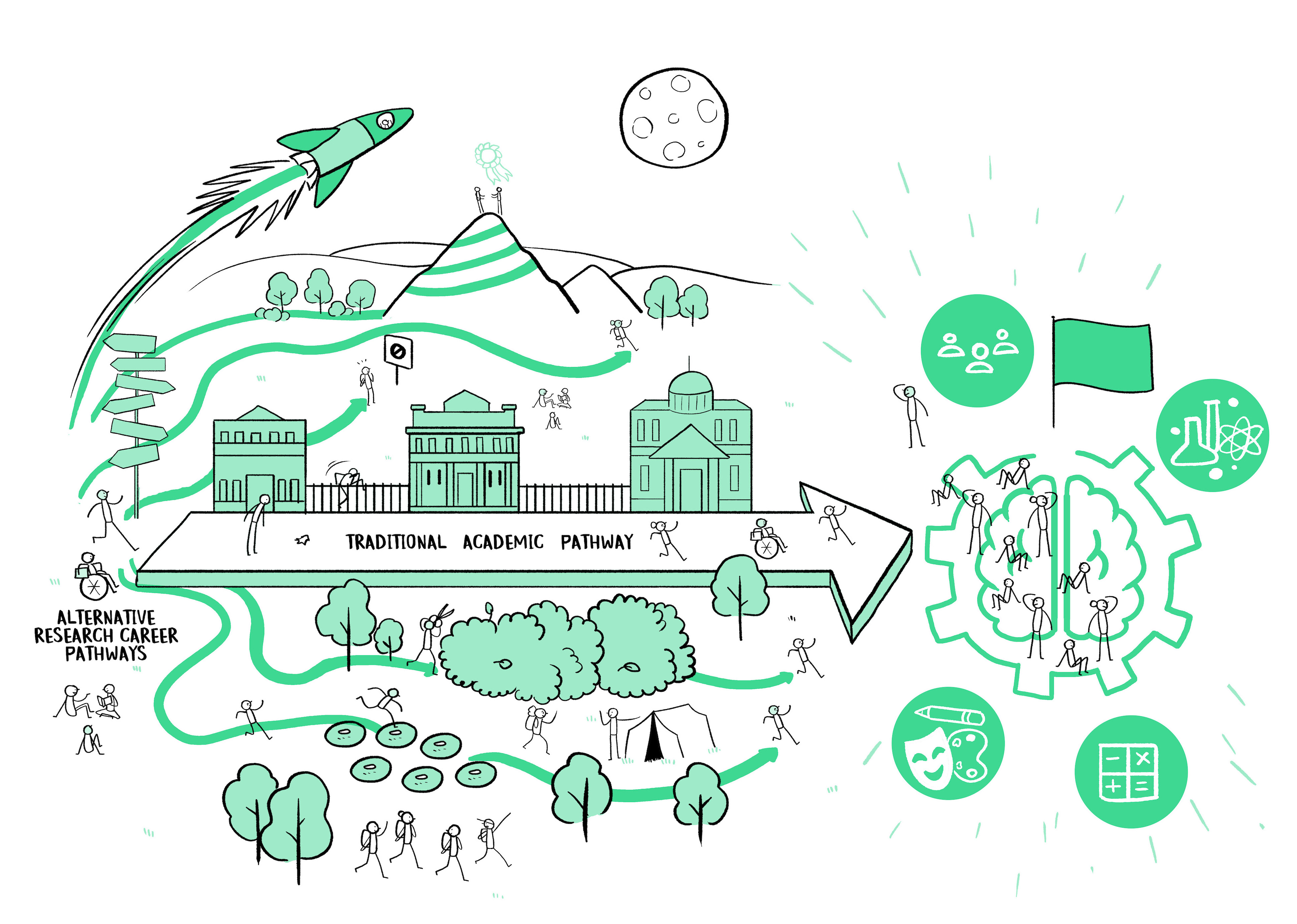}
\caption{\textbf{Navigating professionalisation for diverse roles alongside traditional academic and research roles}: \emph{This figure illustrates 'alternative careers,' often seen on the sidelines of traditional academic or research paths. These diverse sets of emerging skills and roles are known as Research Technical Professional (RTP) or technical professional roles within research and data science organisations. RCM roles, as one of these emerging professionals, play a crucial role in developing and applying innovative approaches to multidisciplinary research, leading meaningful involvement and collaboration among stakeholders, and broadening research impact across sectors.}}
\label{fig:figure4}
\end{figure*}

\subsection{5.1. Articulating Skills and Competencies}

Articulating the RCM skills and competencies is a fundamental first step towards communicating and standardising community management practices and professionalising the RCM roles.
\newline
\newline
In articulating a framework for RCM skills and competencies, as described in this article (section 4.2), we recognise that the definition of research communities is itself broad. A few examples listed in Table 1 represent different types of communities fostered across open research projects (open science, open source), academia and industry partnerships, PPIE projects, research interest groups and community-based participatory research projects. Furthermore, meta-communities that span multiple domains or communities focus on shared practices, standards or policies, such as community for meta-research, meta-practices or professional roles, such as the RCM team, can also be considered a meta-community. The skills required for different RCMs for different projects will therefore vary based on a number of diverse characteristics of their research communities. Strategies for community building will also depend on the scale, size and goals of the project and the intended levels of community participation, as illustrated in Figure 1 and \hyperlink{Sup1}{Supplementary 1}.
Depending on the requirements, the RCM Skills and Competencies Framework can be used both as a reference in creating individual RCM roles and to support their long-term professionalisation in interdisciplinary projects and research organisations.

\subsection{5.2. Aligning Recognition and Reward Systems}

To professionalise RCMs in research, it is essential to proactively evolve standards and criteria for recognition and rewards for incentivising and supporting community management throughout the research process. Historically, academic incentives, including research funding, policy and promotion, have been aligned with the production of traditional research outputs, such as peer-reviewed papers (\cite{Horta_2023}). More attention has been given to writing articles at the end of a research project, often with a senior researcher unilaterally deciding authorship with an outsized focus on first and last authors. This narrow understanding of contributions is increasingly viewed as outdated across academia, and the traditional assessment methods of output are no longer considered fit for purpose (\cite{Brown_2024}). It is important to recognise the work that goes into producing a broad spectrum of tools, documents, partnerships, practices, meetings and insights during the course of research which fail to make it into traditional static publications (\cite{Brandmaier_2024}, \cite{Rushforth_2024}).

\subsubsection{5.2.1 Embedding Considerations for RCMs in Existing Reward Mechanisms}

Rewards for all professions, traditional or emerging roles, are relatively consistent: open acknowledgement and appreciation at work, opportunities for professional growth, performance-based incentives, fair salary and performance-based promotions or exposure to new skills through training opportunities (\cite{Dewi_2022}, \cite{Garbers_2014}, \cite{Luthans_2000}). RCMs can be fairly rewarded through these existing mechanisms by evaluating their success appropriately in line with their professional responsibilities. When roles are professionalised, employees develop stronger agency and motivation in their positions, while institutions can better recognise and reward their contributions to the progress and success of research (\cite{Rai_2018}, \cite{Shafiq_2009}). In the case of RCMs, professionalisation can enable institutions to fulfil their responsibility to engage and support collaboration among the various research and non-research groups involved, including the public (\cite{Santos_2021}, \cite{Sorensen_2021}).
\newline
\newline
To fully realise the potential of community management, all stakeholders in research should view community building as a shared responsibility and a crucial objective. In the study on material and social contexts of research, authors (\cite{Leonelli_2015}) concluded that the “emergence of a resilient research community is partly determined by the degree of attention and care devoted by researchers to material and social elements beyond the specific research questions under consideration”. To ensure long-term collaboration and the benefit of research partnerships and community engagement, institutions must, therefore, support and incentivise all members of a research community to actively engage with their community efforts. In a research environment where community management is regarded as a critical and respected function, research communities can thrive and sustain both their individual and collective activities (\cite{Young_2013}). Professionalisation of RCMs should therefore be seen as steps towards fostering a research culture that values, supports, and rewards the contributions of community members in terms of both technical expertise and social experiences. 

\subsubsection{5.2.2 Contributing to and Leveraging Research Assessment Frameworks}

Initiatives such as the DORA (\cite{DORA_2024}), CRediT (\cite{Holcombe_2019}), the Royal Society's Resume for Researchers (\cite{RoyalSociety_2024}), the Dutch Recognition\&Rewards programme \cite{RecognitionRewards_2024}, COARA \cite{CoARA_2024} and HiddenREF (\cite{HiddenREF_2024}) campaigns, are all seeking to build a new consensus around a broader scope for research contributions. Community-driven projects like The Turing Way (\cite{TuringWay_2024b}), African Open Science Platform (\cite{AOSP_2024}), EOSC Association Task Forces (\cite{EOSC_2024}), Association of Research Managers and Administrators - UK’s ARMA and European counterpart EARMA (\cite{EARMA_2024}), RSE communities (\cite{SocRSE_2024}), Technician Commitment (\cite{TechniciansCommitment_2024}), PRISM network (Professional Research Investment and Strategy Manager) (\cite{PRISM_2024}) and ResearchOps (\cite{ResearchOps101_2024}), among many other international initiatives, are increasingly influencing funders and policymakers to evolve their metrics for evaluating academic success.
\newline
\newline
This shifting narrative is an opportunity to account for the wide range of interdisciplinary roles and research communities in shaping research processes, outcomes and their impact. The professionalisation of RCM roles must occur alongside these initiatives by advocating for and informing the evolving metrics of research assessment with the requirements for community building in research. RCM roles focus on the research process, from project design to the production and maintenance of different kinds of outputs, consideration for which should be a central tenet to aligning assessment systems with their professional recognition. Therefore, the recognition effort should focus on elevating RCMs’ mission to embed best practices in research and catalyse connections across communities.

\subsubsection{5.2.3 Aligning with National and International Skills Priorities}

Investments from funders nationally and internationally are key to improving the status and professional security of community professionals like RCMs and other RTPs or infrastructure roles. For example, following national strategy and policy recommendations in the UK, UK Research and Innovation and EPSRC announced £16 million to support research technical professionals in community and capacity-building efforts in 2023 (\cite{UKRIEPSRC_2024}). This is one of several funding investments in digital research infrastructure professionals including community developers (RCMs), data wranglers, technicians and RSEs. One such project is People in Data (Table~\ref{Table1}, project 10), a community-based project funded by EPSRC and co-led by a member of the RCM team, which will convene existing communities to share knowledge, address critical challenges faced by different stakeholders working with data and create wider visibility and adoption of data-focused skills. Several such initiatives will be designed over the next few years, which will require expertise from infrastructure roles with specific opportunities to involve RCMs in leadership positions equipping them to engage all stakeholders of a data ecosystem in addressing shared issues around data science and AI skills in different domains and sectors.
\newline
\newline
Similarly, philanthropic and public sector investments are being made to establish Open Source Program Offices (OSPOs) within universities and public sector organisations (\cite{Munir_2021}). 
Although the open source movement has been around for nearly three decades, OSPOs are a relatively new phenomenon, originally emerging from the tech industries in the USA and European countries (\cite{Ruff_2022}). An OSPO team requires experts like RCMs dedicated to facilitating community engagement efforts to support community collaboration, skill building, and the adoption of open source practices to drive innovation and greater return on investment across the organisation. Rather than a few specific projects, OSPOs aim to improve the general use, development practices and reusability of open source software, an area or work that has clear overlap with community building and RCM roles. Since 2023, the United Nations has hosted an international symposium on OSPOs for Good to enable knowledge sharing among stakeholders from government, public sector, industry and academia from across the globe (\cite{UN_2024}). 
\newline
\newline
Given the increased awareness and potential for OSPOs to promote open source practices for developing digital public infrastructure and resources, RCM and infrastructure roles like RSE and RAMs can strategically combine their expertise and support open source-led innovative solutions embedding shared practices and collectively contributing to OSPO’s goals at national and international levels (\cite{Jimenez_2021}).
\newline
\newline
The goals of research assessment, policy and funding is to improve desired skills at all levels of research, future-proof the talent pipeline at national and international levels and unlock the economic potential and societal benefits of RRI. RCMs’ involvement from across different disciplines, organisations and countries in achieving these goals by engaging different professionals and involving community stakeholders clearly demonstrates the value of professional RCMs for the research ecosystem and the wider economy.

\subsection{5.3 Building Long-Term Career Pathways}

\begin{figure*}
\centering
\includegraphics[width=1\textwidth]{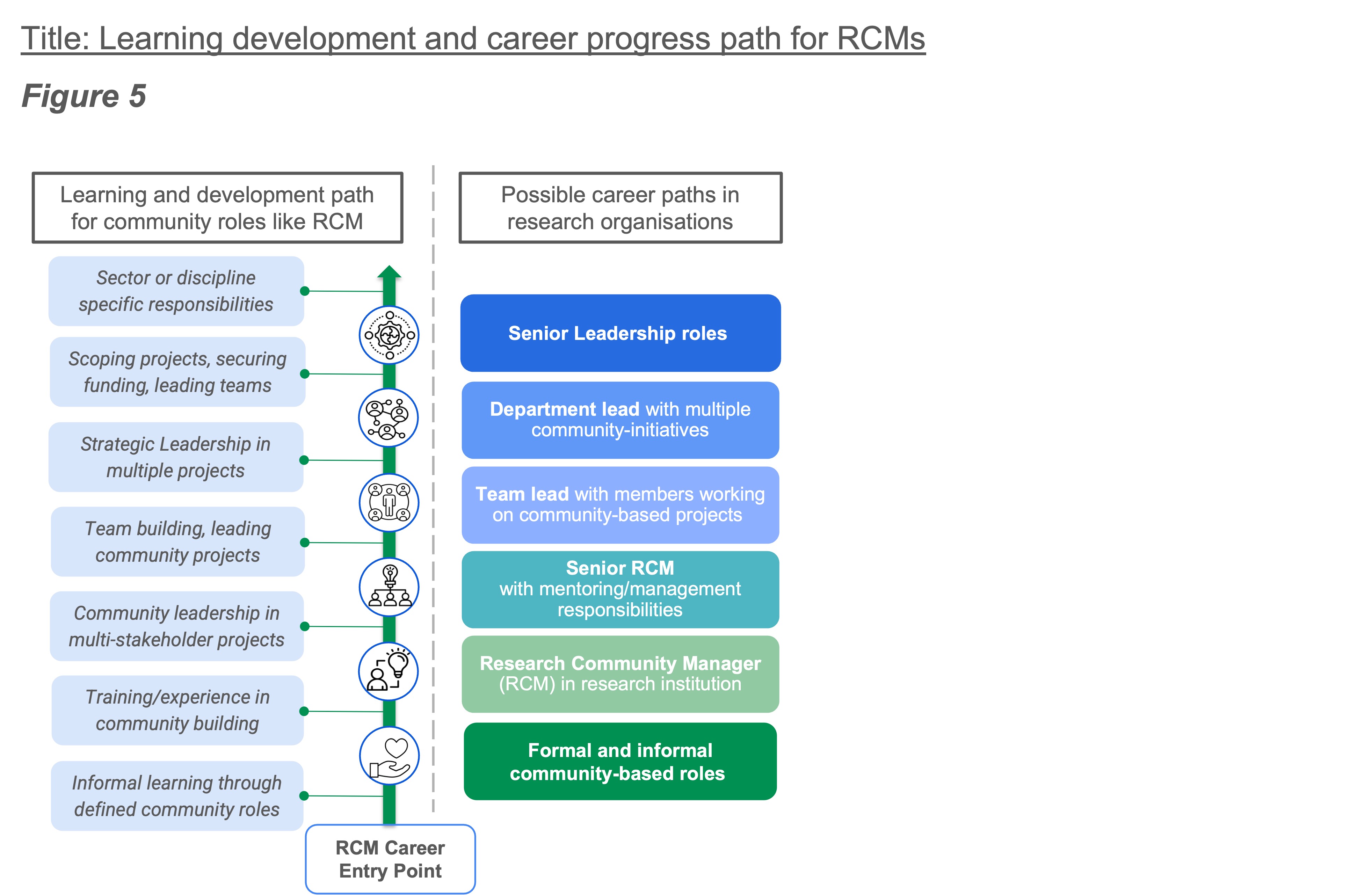}
\caption{\textbf{Learning development and career progress path for RCMs}:
\emph{This figure illustrates the learning and development (left panel) and career progression (right panel for RCM. An individual can start their career journey in community management through informal learning in roles such as conference organising committee members and volunteer leadership in community projects. Progression includes formal training and experience through dedicated roles like RCM within research organisations, comparable to academic researcher and junior community manager roles in academia and industry respectively. An RCM can advance to a Senior RCM role through leadership in multi-stakeholder projects, akin to postdoctoral and senior management positions. Further advancement can lead to RCM Team Lead which can involve managing multiple projects and teams, similar to principal investigator and team lead roles. Ultimately, RCMs can become department leads, overseeing multiple teams, securing funding, and leading strategic initiatives. An extended version of this image, illustrating career progression in community management roles in academia and industry is provided in Supplementary 4}.}
\label{fig:figure5}
\end{figure*}

RCMs bring practical experience in community building that they often learn as part of their previous roles, often making a career switch from traditional research roles to focus on community management. Several organisations offer learning resources as well as paid certification for community managers. Nonetheless, this route is far from widespread or suitable practices for most research institutions. These roles have largely remained informal, often filled by researchers who are self-taught through responsibilities such as managing scientific networks or supporting open source communities (Figure~\ref{fig:figure5}).
\newline
\newline
RCMs transfer their knowledge about how research and data science projects work and the community management practices that they apply at different stages to enhance the impact of their research. Previous research and industry experience allow RCMs to apply leadership skills such as taking initiative as an independent researcher, proactively involving or mentoring community members, recognising challenges and opportunities, and proposing solutions in time to support their communities. Therefore, formal RCM positions should not be seen as stepping away from research but as stepping in from another perspective, to gain valuable additional and specialist research skills and deepen their leadership skills to progress further in their research career. 

\subsubsection{5.3.1 Community Champion Roles}

To fulfil the foundational need of community management, researchers interested in RCM roles bring some experience in organising communities of different sizes, natures and formats in different capacities. For example, managing a journal club, maintaining an open source tool or coordinating a scientific network. Researchers can then choose to specialise in community management through learning and practical experience, for example, taking community management training, becoming a core maintainer of a public resource or leading the communication processes for their research network.
\newline
\newline
These researchers make ideal candidates for RCM roles in a well-defined project-based CoP involving a limited number of partnering organisations or teams participating in well-scoped community-based activities. These RCMs act as community champions with clear operational responsibilities and opportunities to grow in their roles. Irrespective of their responsibility types or decision-making authority, RCMs apply open leadership skills in their roles in leading and nurturing their communities. Ideally, RCMs should have access to appropriate levels of supervision, for example, regular mentoring and support should be provided to those who don't have previous experience in professionally managing research communities.

\subsubsection{5.3.2 Strategic roles for multi-stakeholder and complex communities}

With more experience, RCMs assume more advanced or independent leadership roles in their communities. An RCM can progress in their careers to take senior RCM roles working with more complex communities such as managing stakeholders at a department level or coordinating collaborations among multiple research groups and institutions. 
They can also provide mentorship and guidance to other community members or formally manage other RCMs. They design training programs focused on skills needed in their communities, offer consultation on community strategies, lead initiatives, and report on community progress, and engage in internal advocacy efforts. This often involves working more closely with senior leaders and therefore at a strategic level, providing direct leadership for the wider community. This progression emphasises continuous learning, strategic thinking, and the ability of RCMs to support others within the community management domain.

\subsubsection{5.3.3 RCM's role as the stepping stone for future leaders}

Gaining more experience in various community roles is an excellent building block for progressing to senior leadership roles. As community leaders, experienced RCMs build a strong identity and knowledge in research and best practices, as well as bring a deep understanding of how collaborations in multiple contexts affect the project development, team's growth and research culture, as well as its impact in the broader society.
\newline
\newline
With experience, RCMs bring more strategic and leadership approaches into their work, expand their professional network and gain hands-on experience in dealing with new challenges. Increasingly, community management skills and expertise are becoming important for all kinds of leadership roles (\cite{Crawford_2006}, \cite{TCR_2020}). With their broad range of expertise, an RCM's career progression could be in academia as a senior RCM, senior researcher, team lead, or principal investigator role and programme director (such as the TPS programme at the Turing) working to advance community management or community-based research in their respective domains.
\newline
\newline
As with any professional, it is important to recognise that the career goals for each RCM will differ based on their skill sets and personal interests. RCMs may stay in their roles or decide to move to different sectors, like industry, non-profit or independent research institutions. They may also assume roles like lead of community programmes, department head or director of a community organisation. With their exposure to a wide range of strategic and operational responsibilities, they can also switch job types or pursue consultancy careers. By stretching beyond the strict remit of research, silo-prone public sectors and administration are also examples of settings or professional environments that could strongly benefit from RCM expertise in their goals to improve public engagement in decision-making and public trust through participatory approaches (\cite{Scher_2023}).
\newline
\newline
Based on their career aspirations, RCMs can prioritise what kinds of skills they strengthen in their community roles, how they grow in their profession and which career paths they take for themselves. Regardless of their future directions, RCMs, like in any secure job, should be given a good working environment, stable contract conditions and appropriate compensation with professional benefits.

\subsection{5.4 Offering Prolonged Mentorship and Support}

As highly connected members, RCMs often become direct points of contact for community members and frequently respond to their queries. They address various community requests, many involving repetitive tasks or context-switching throughout the day, such as offering onboarding and upskilling opportunities to different members at different timescales. At the same time, RCMs are responsible for engaging the community in ongoing work, producing community documentation and strategically updating their approaches across different stages of the research lifecycle in response to the maturation and evolving needs of the community. These responsibilities can significantly increase RCMs’ workload, making it a demanding and potentially stressful occupation (\cite{deWinter_2017}), especially if RCMs are not provided with appropriate professional support to manage them. Therefore, long-term mentoring and institutional investment should be provided to effectively support RCMs in their roles.

\subsubsection{5.4.1 Mentoring RCMs}

Even the most experienced RCMs have to continuously improve their methods while keeping up with the changing requirements and trends of data science and communities. Community management training and resources can be offered at the initial stages of professional development, however, at the later stage, RCMs have to be mentored in identifying, learning and adapting new tools, practices and platforms for community management. Therefore, on an ongoing basis, an RCM should have access to advisory support for situation-specific strategic and management guidance from other experienced professionals. 
Peer-to-peer engagement with other RCMs can be particularly beneficial for the professional growth of individual RCMs. Meta-communities like interest groups, community forums and networking events especially focused on RCMs and other infrastructure roles provide a good source for informal learning, along with peer-to-peer learning within teams of RCMs. For example, many open source projects that manage large volunteer communities or dedicated networks of community professionals, provide informal support and opportunities for skill development around community management. Some non-profit and consulting organisations also offer targeted training, mentoring and coaching for individuals and teams supporting scientific communities (\cite{OLS_2024}, \cite{Openscapes_2024}). These spaces provide opportunities for peer-based learning through targeted collaboration, theme-based knowledge building and chance interactions with other individuals on shared and complementary skills. These can often become avenues for creative discussions and knowledge exchange with fellow community managers, enabling RCMs to gain second-hand experience on different challenges and identify appropriate approaches for their contexts.

\subsubsection{5.4.2 Institutional Support Network for RCMs}

Although external communities and networks of like-minded people become crucial for personal development, they often don’t respond to organisational complexity. There needs to be a more systematic solution for support for professionalising RCM roles within an organisation. In a traditional research team, senior members and leaders of projects are expected to offer general leadership guidance and advice, but in the case of RCMs, they may not always have the time, expertise or resources to professionally support their community management work. This can lead to individual RCMs working alone or feeling isolated when advocating for the community's interests in their projects by themselves. This is where a dedicated RCM team can provide a more sustainable infrastructure for the professional development of RCM roles within the institution.
\newline
\newline
An organisationally-supported RCM team offers structured mentorship, training and hands-on support to all RCMs at the organisation. As shared in the example from the Turing, the RCM team manages, supports and coordinates with other RCMs working on different projects. They apply general team management approaches, such as team onboarding, training and regular meetings to keep each other updated and offer ad-hoc support. They also engage in more targeted activities such as team development activities, shadowing and a regular reporting process to improve skills and support for RCMs. They offer need-based consultation on other projects, participate in strategic development discussions and communicate about the team's work at institutional or external platforms to establish and advocate for their professional identities. We discuss this aspect in detail in the next section.

\section{6 An RCM Team Can Respond to Institutional Barriers to Community Building}

RCM roles are often designed to facilitate collaborations with different stakeholders within and beyond the project as needed. To engage different specialist groups from beyond the project team, they build a good understanding of what different departments and specialist groups at an organisation do, how they operate, where they should be involved and what expertise can be invited at different stages of a project. 
Despite their efforts to connect and engage across an organisation, there are systemic challenges that can pose barriers and restrict the success of the RCM roles, some of which include:

\begin{itemize}
    \item It is within the scope of RCM's responsibilities to facilitate collaboration between the project team and other specialist groups including external stakeholders they can work with in addressing a research question. However, it is impossible for one person alone - even a skilled RCM - to build comprehensive knowledge about different specialist groups unless all members of the project participate in community building (\cite{Golenko_2012}, \cite{Santos_2021}). 
    \item When the right collaborators are identified, RCMs are tasked with onboarding and actively involving them in the project. However, if researchers do not see direct values of open collaboration or lack the capacity to engage with new collaborators in the broader scope of the project, it can pose a major challenge for RCMs in achieving their goals in the project \cite{Brocke_2015}, \cite{Delgadillo_2016}.
    \item A project team may be fully supportive of RCM work, but if they lack buy-in from institutional leadership and therefore other stakeholders of the institute, RCMs’ efforts may go unrecognised, unsupported or not credited appropriately for the project's success (\cite{Smith_2022}).
    \item Arbitrary classification of RCMs as non-research or ‘research support from the sidelines’ can result in their exclusion from discussions relating to project strategy or decision-making that impacts the broader community. This can prevent RCMs from advocating for community interests and lead to overlooked opportunities in research that effective community management could otherwise bring to the project (\cite{Macaulay_1999}).
\end{itemize}

Such challenges can lead to underutilisation of RCMs’ professional expertise, leaving them unmotivated, excluded and burned out. Addressing these organisational challenges is, therefore, crucial for ensuring that RCM roles are successful in their roles. An institutionally supported RCM team can become an organisational response and systemic solution for navigating organisational complexity in building research communities (\cite{Leonelli_2015}, \cite{Vasileiadou_2012}).

\subsection{6.1 Advantages of RCM Teams as part of the Professionalisation of Community Management}

A key advantage of operating as a team over isolated RCM roles lies in the ability to facilitate cross-community collaboration, retaining and enhancing institutional knowledge from across different projects. As a well-connected group, RCMs are also able to extend existing research-based solutions and partnerships from one context to another. Leveraging the existing knowledge of experienced members, RCMs can readily identify people and teams from beyond their respective projects, build on existing partnerships and engage with specialists on specific objectives.
\newline
\newline
As part of a research project, RCMs are responsible for project-specific CoPs, and as a part of the RCM team they provide leadership in bringing alignments between different project-specific CoPs at the organisation level. By strengthening connections between different efforts, the RCM team promotes cross-exchange of reproducible and generalisable research outputs that may otherwise remain limited to individual projects. Members of an RCM team collectively maintain and re-use community resources, standardised practices and infrastructures that can be adapted across different projects.
In the context of professionalisation, RCM teams exhibit institutional commitments to both building research communities and offering professional support for RCMs through prolonged mentoring, coaching and peer-to-peer support. RCMs in collaboration with other team members build standard approaches for community management across the organisation and improve efficiency by reusing existing resources, maximising the return on investment already made by the organisation. RCM teams improve accessibility, benefits, and impact of project outcomes, as well as facilitate continuous learning for all stakeholders. They transfer knowledge between projects, both scientifically and socially, facilitating new collaborations and interdisciplinary research ideas between groups that may otherwise depend on chanced interactions.
\newline
\newline
Mentoring new RCMs, and eventually managing a team of RCMs is a natural career progression for experienced RCMs, such as in the case of the Turing’s Senior RCMs and leads of the RCM team (Figure 5). Senior members of the team are involved in recruiting, onboarding and training new members who may bring varying levels of domain expertise and different community management approaches. They guide other RCMs in their day-to-day tasks and support the team in exchanging resources and developing best practices at the institutional level. They also bring systems-level understanding, passing down institutional knowledge from their work and connecting RCMs to other opportunities that can help them gain new skills and recognition within the organisation.
Beyond project-based obligations, members of the RCM team support and maintain theoretical and practical knowledge within a project in the form of shared documentation, tools, methods and technical or non-technical reports. Their leadership in sharing and dissemination of these resources through central repositories, open communication, publications and advocacy through professional networks should be supported and recognised like any other research output.

\subsection{6.2 Routes to Formalising RCM Roles in Resource-Constrained Organisations}

We recognise that cost constraints may hinder the hiring of dedicated RCMs in some organisations, especially in universities with limited resources and research institutions with restricted investments in emerging opportunities (\cite{Trevors_2010}). However, research organisations can take various approaches to recognise and encourage community management skills, contributing to professionalisation efforts even when new RCM roles cannot be formalised. For example, using the RCM Skills and Competencies Framework (described in section 4.2), research teams can incorporate RCM skills and responsibilities into existing job descriptions for researchers. Researchers in a project can also share RCM responsibilities as part of their role within a project. Especially for projects that are interdisciplinary or should involve multiple stakeholders, different roles within a research team can integrate best practices from across core and peripheral aspects of RCM work even when RCM roles can’t be formally allocated.
\newline
\newline
In the absence of institutional capacity to support teams of full-time RCMs, we discourage RCM responsibilities from being delegated only to a few willing researchers without appropriate recognition and compensation, or consideration for their career development (\cite{Smithers_2022}). Institutions can provide mechanisms for researchers with an interest in community management to gain community management experience as part of a CoP such as an interest group. Participation in these spaces can be supported through paid roles such as project-based fellowship opportunities or through skill development support through professional training. Organisations can also create part-time or short-term community management roles. However, these roles should be designed with defined goals, for example, the development of non-traditional research outputs, implementing open source and reproducibility practices or specific community engagement opportunities.

\subsection{6.3 Adopting Approaches for the RCM Profession from the RSE Movement}

The RSE profession offers a strong model of professionalisation of emerging roles in data science (\cite{Katz_2023}, \cite{Sims_2021}, \cite{Woolston_2022}). As an emerging profession with relatively new ways of working in socio-technical contexts of research and data science, RCMs can build on the approaches that have led to the success of the RSE movement. RCM teams can already draw enormously from established teams of RSEs, such as the Turing’s Research Engineering Group. They can also combine their efforts with the RSE community to exchange practices for the professionalisation of new RTPs and infrastructure roles in interdisciplinary research (\cite{Sharan_2024}, \cite{UKRISTP_2024}). In addition, research-based technical approaches and evolving practices of community management should be supported by organisations, funders and policymakers to support healthy, resilient and thriving CoPs instrumental for strengthening RRI nationally and internationally (\cite{Hoeven_2013}, \cite{Owen_2012}).
\newline
\newline
In addition, RCMs should be supported as researchers in their own profession and practices contributing to the advancement of this profession. For example, members of the RCM team should be encouraged to engage with the national and international initiatives involved in the recognition and professionalisation of data science roles. In collaboration with other RTPs, RCMs can contribute to building sustainable resources, promote open source practices, and enhance training and career pathways, creating a more integrated and supportive environment for RTPs and infrastructure roles as part of team science.
\newline
\newline
Turing’s RCM team members contribute to initiatives such as HiddenREF, People in Data (UKRI and EPSRC funded RTP project), UK Reproducibility Network, OSPOs for Good and Society of RSE, where they present their work advocating for RCM roles alongside other RTPs. Building on the momentum of the RSE movement, the Turing’s RCM team leads have also been researching the professionalisation of diverse data science roles (Table 1, project 5) and developing resources, including this article, for better recognition of the Research Community Management as an important data science profession (\cite{Karoune_2024}, \cite{Sharan_2024}).

\section{7 Conclusion}

Establishing RCM roles as a viable career path requires recognising community management as a dedicated occupation distinct from part-time or side jobs expected to be fulfilled by volunteer positions. RCMs take a broad spectrum of responsibilities in research teams and embed a diverse range of skills and competencies that part-time or incidental roles cannot fulfil. Research teams and organisations should, therefore, hold responsibility for prioritising community-related objectives to improve collaboration in their research. RCMs can lead on these objectives, while being fairly recognised, supported and compensated for their work in engaging research stakeholders including the community of users and beneficiaries appropriately with different aspects of research. Well-defined RCM roles as a professional option will broaden career opportunities for researchers including PhD students and postdocs beyond the traditional academic path.
\newline
This article offers a Community Maturation Indicator and an RCM Skills and Competencies Framework for community management roles, conceptualised for research and data science organisations. These frameworks can be used for creating RCM roles and incorporating community management expertise in multi-stakeholder research projects and interdisciplinary initiatives on research and data science. In addition, we proposed a roadmap for professionalising RCM roles by building a shared understanding of their skills and competencies, recognition and rewards, career pathways and support infrastructure. 
\newline
\newline
Since the focus of RCM roles is different from traditional research roles, their work should not be measured in terms of traditional research activities, such as producing research articles that are typically used to assess a project's success, but across different areas of research as discussed in this article. Alongside policies and initiatives supporting the improvement and reformation of academic incentives and rewards, this article aims to influence research assessment frameworks, so that different kinds of contributions to research, such as community building, are considered important for research career progression. The slow evolution of research assessment metrics and outdated measures of success should not hinder the professional recognition of RCMs and their career advancement.
\newline
\newline
In this article, we have shared insights from The Alan Turing Institute, a national institute in the UK that leverages its convening role to foster communities of researchers, practitioners and decision-makers to navigate and address complex societal challenges with data science and AI. At the Turing, RCMs work alongside traditional research roles, such as project leads, postdoc and doctoral researchers and RTPs or infrastructure roles like RSEs, Data Wranglers and RAMs that constitute the institute's core capabilities involved in the development of data science and AI solutions responsibly in the real-world contexts. With different frameworks and case studies from the Turing’s RCM team, we provide a working example of how institutions can integrate community management expertise across different projects and incentivise community collaboration as an important part of research goals. Evidence-based resources, such as this article, sharing insights from organisationally supported RCM teams will contribute to the broader professionalisation efforts of RTPs and infrastructure roles, especially in the contexts where RCM roles have yet to be formalised.
\newline
\newline
With examples from the Turing, we strongly recommend that institutions invest in establishing RCM teams and formalising individual RCM roles. RCM teams provide RCMs with early feedback on their work, peer mentoring to support their professional development and share collaborative opportunities that they may otherwise not have in their specific project. As institutionally supported research groups, RCM teams can systematically dismantle the silos that traditionally exist in academic research, fostering more connected research teams and maximising the impact of their outcomes.

\section{Authors Contribution Statement}
\begin{itemize}
    \item Conceptualisation: MS, EK, AL, KW
    \item Funding acquisition: KW, MS, EK, AB
    \item Supervision: KW, MS, EK
    \item Project Administration: KW, MS, EK, AB, AAA
    \item Writing – Original Draft Preparation: MS, EK
    \item Writing – Case studies from the Turing: VH, GK, AAA, SB, ALS, MS, EK
    \item Writing – Original Draft Review: MS, EK, VH, CGVP, GK, AB, KW
    \item Writing – Original Draft Editing: MS, KW
\end{itemize}
 
\section{Acknowledgement}
We, the authors of this manuscript, would like to express our sincere gratitude to the Tools, Practices, and Systems team, especially the Research Application Managers, Senior Researchers, and Research Project Managers for their invaluable support to the Research Community Management team. We also thank the Turing project teams, research staff and collaborators for their support and engagement with our team.
\newline
We would like to thank Professor Rachel Hilliam for her discussions, encouragement, and valuable feedback on earlier versions of this paper as part of the Turing’s Skills Policy project, 'Professionalising Traditional and Infrastructure Research Roles in Data Science'. We express our gratitude to The Turing’s Skills team for their support in developing the related research outputs funded by the Skills Policy Award between 2023 and 2024.
\newline
Finally, we extend our thanks to The Turing Way community and the members of the communities supported by the RCMs, whose contributions to various CoPs and engagement with community management have been essential to the development of the RCM roles.

\section{Declarations of Conflict of Interest}

The authors declared no potential conflicts of interest with respect to the research, authorship, and/or publication of this article.

\section{Funding}

EK and MS’ work on this paper was supported by the Alan Turing Institute’s Skills Policy Award, supported by the Ecosystem Leadership Award under the EPSRC Grant EP/X03870X/1.
\newline
This work was also initiated through Wave 1 of The UKRI Strategic Priorities Fund under the EPSRC Grant EP/T001569/1 and EPSRC Grant EP/W006022/1, particularly the “The Tools, Practices and Systems" theme” within those grants. All authors acknowledge support from the Alan Turing Institute.
\newline
\newline
Members of the Research Community Management Team work on multiple data science and AI projects at the Alan Turing Institute, and therefore, we would also like to acknowledge the following funding sources that fund their positions:
\begin{itemize}
    \item MS and AAA’s roles in The Turing Way Practitioners Hub and BridgeAI are funded by Innovate UK BridgeAI. This project has also received funding and support from the Ecosystem Leadership Award under the EPSRC Grant EP/X03870X/1. KW is funded as the principal investigator for BridgeAI.
    \item EK was an SRCM for the Turing-RSS Health Data Lab, funded by The Department for Health and Social Care (Grant Ref: 2020/045) with in-kind support from The Alan Turing Institute and The Royal Statistical Society.
    \item VH is an RCM for the Turing-Roche Strategic Partnership; funded by Roche through a five-year strategic investment in this research partnership. 
    \item GK is an SRCM for Data-Centric Engineering Programme; funded under the Lloyd’s Register Foundation grant G0095.
    \item CGVP is an SRCM for Environment and Sustainability Grand Challenge, funded by the Engineering and Physical Sciences Research Council (Grant number EP/Y028880/1) 
    \item EK and SB's roles are funded by the NIHR Artificial Intelligence for Multiple Long-Term Conditions (AIM) Research Support Facility (NIHR202647). The views expressed are those of the authors and not necessarily those of the NIHR or the Department of Health and Social Care. KW is funded as the principal investigator for the AIM RSF.
\end{itemize}

\bibliography{references}

\section{Supplementary Files}

\emph{Following Supplementary Tables and Figure have been provided, which start on the next page.}

\begin{itemize}
    \item \hyperlink{Sup1}{Supplementary 1} (Table)
    \item \hyperlink{Sup2}{Supplementary 2}  (Table)
    \item \hyperlink{Sup3-1a}{Supplementary 3.1}  (Table)
    \item \hyperlink{Sup3-2}{Supplementary 3.2}  (Table)
    \item \hyperlink{Sup4}{Supplementary 4} {Figure}
\end{itemize}


\begin{table*}[ht]
\caption*{\textbf{Supplementary 1}}
\hypertarget{Sup1}{}

\setlength{\tabcolsep}{3pt} 
\renewcommand{\arraystretch}{1.5} 
\begin{tabular}{p{0.12\linewidth} | p{0.12\linewidth} | p{0.12\linewidth} | p{0.12\linewidth} | p{0.12\linewidth} | p{0.12\linewidth} | p{0.12\linewidth}}
\toprule
\"Columns: Levels of Community Participation" / "Rows: Stages of Community Building"& Level 1: Inform Community & Level 2: Invite Feedback	& Level 3: Engage and Involve	& Level 4: Mobilise and Connect	& Level 5: Empower Groups	& Level 6: Decentralise Power\\ \hline
\midrule
    \texttt{Stage 1: Initiation} &	Information about the project	&	Community /stakeholder awareness plan	&	Knowledge share activities	&	Open communication strategy	&	User engagement data	&	Reusable resources for community members \\
    \texttt{Stage 2: Planning and Design}	&	Project roadmap created by project team	&	Community /stakeholder  feedback process	&	Feedback activities	&	Feedback on ongoing and future plans	&	Feedback and community interest data	&	Onboarding new teams and members \\
    \texttt{Stage 3: Implementation} &	Roadmap created with collaborators 	&	Community /stakeholder  engagement process	& Active engagement activities	&	Involving the community in delivering project objectives	&	Feedback and community impact data	&	Creating new leadership opportunities \\
    \texttt{Stage 4: Growth and Scaling}	&	Roadmap created with community groups 	&	Community /stakeholder collaboration process	&	Collaborative process for group engagement	&	Mobilising community-led initiatives in line with project objectives	&	Reporting and Impact data from community initiatives	&	Groups connected and led towards shared goals \\
    \texttt{Stage 5: Monitoring and Evaluation}	&	Roadmap draft created and revised by community groups	&	Community /stakeholder involvement in decision-making	&	Multiple collaborative group activities	&	Community groups involved in decision-making	&	Reporting and impact data from community initiatives	&	Connected groups shaping future goals and directions \\
    \texttt{Stage 6: Sustainability or Sunsetting} &	Roadmap draft created and iteratively updated by community	&	Community-led decision-making process	&	Activities to delegate responsibilities to groups	&	Community groups making decisions in benefit of the community	&	Independent reporting from connected groups	&	Connected groups working independently and collectively with agency \\
\bottomrule
\end{tabular}\\[10pt]
\begin{center}
      \emph{\textbf{Supplementary Table 1: Community Maturation Indicator}. This table provides the raw data used in Figure 1b of the article, showcasing an example of a community-building strategy that can be applied at both Stage 1 and Stage 6 across all Levels of Participation. Additionally, it includes examples of community-building strategies for all other Maturation Statuses, spanning Stages 2-5 in the community-building process for all Levels of Community Participation.} 
\end{center}
\end{table*}


\begin{table*}
\small\sf\centering
\caption*{\textbf{Supplementary 2}}
\hypertarget{Sup2}{}

\setlength{\tabcolsep}{3pt} 
\renewcommand{\arraystretch}{1.5} 
\begin{tabular}{p{0.02\linewidth} | p{0.15\linewidth} | p{0.05\linewidth} | p{0.1\linewidth} | p{0.2\linewidth} | p{0.1\linewidth} | p{0.2\linewidth}}
\toprule
\#&Project name&Project Start Year&RCM start year and current status&Community Status on the Community Maturation Indicators&Relevant links&Manuscript authors working on this project\\ \hline
\midrule
    \texttt{1} & \emph{The Turing Way} & 2019 & 2019 - Ongoing & Empower Groups \textbf{(Level 5)} - Monitoring \& Evaluation \textbf{(Stage 5)} & \href{https://www.turing.ac.uk/research/research-projects/turing-way}{\emph{Details}}, \href{https://github.com/alan-turing-institute/open-research-community-management/blob/main/team-related-comms/briefing-notes/2022-ASG-Briefing_0_TuringWay.pdf}{\emph{Briefing Note}}, \href{https://github.com/the-turing-way/}{\emph{Repository}} & 2022-Present: Anne Lee Steele (ALS), 2020-2021: Malvika Sharan (MS), Kirstie Whitaker (KW), Arielle Bennett (AB), Previous RPM Alexandra Araujo Alvarez (AAA) (2022-2023) \\
    \texttt{2} & The Turing-Roche Strategic Partnership & 2021 & 2021 - Ongoing & Engage and Involve Members \textbf{(Level 3)} - Growth and Scaling \textbf{(Stage 4)} & \href{https://www.turing.ac.uk/research/research-projects/alan-turing-institute-roche-strategic-partnership}{\emph{Details}}, \href{https://github.com/alan-turing-institute/open-research-community-management/blob/main/team-related-comms/briefing-notes/2023-RCM-Briefing_Turing-Roche_community-VH.pdf}{\emph{Briefing Note}}, \href{https://github.com/turing-roche}{\emph{Repository}} & 2021-Present: Vicky Hellon (VH)\\
    \texttt{3} & Turing-RSS Health Data Lab & 2021 & 2021 - Concluded in 2023 & Engage and Involve Members \textbf{(Level 3)} - Implementation \textbf{(Stage 3)} & \href{https://www.turing.ac.uk/research/research-projects/turing-rss-health-data-lab}{\emph{Details}}, \href{https://github.com/alan-turing-institute/open-research-community-management/blob/main/team-related-comms/briefing-notes/2023-RCM-Briefing_Turing-RSSHealthDataLab_community-EK.pdf}{\emph{Briefing Note}} & 2021-2023: Emma Karoune (EK) \\
    \texttt{4} & AI For Multiple Long-term Conditions -Research Support Facility (AIM-RSF) & 2021 & 2021 - Ongoing & Invite Community Feedback \textbf{(Level 2)} - Implementation \textbf{(Stage 3)} & \href{https://www.turing.ac.uk/research/research-projects/ai-multiple-long-term-conditions-research-support-facility}{\emph{Details}}, \href{https://github.com/alan-turing-institute/open-research-community-management/blob/main/team-related-comms/briefing-notes/2023-RCM-Briefing_AIM-RSF-OpenCollab_community-EZ.pdf}{\emph{Briefing Note}}, \href{https://github.com/aim-rsf}{\emph{Repository}} & 2021-Present: Sophia Batchelor (SB), 2024-Present: Emma Karoune (EK), 2021-2023: Eirini Zormpa, Kirstie Whitaker (KW) \\
    \texttt{5} & Professionalising Data Science Roles - Turing's Skills Policy Award & 2023 & 2023 - Concluded in 2024 & Invite Community Feedback \textbf{(Level 2)} - Implementation \textbf{(Stage 3)} & \href{https://github.com/alan-turing-institute/professionalising-data-science-roles}{\emph{Details}}, \href{https://github.com/alan-turing-institute/professionalising-data-science-roles}{\emph{Repository}} & 2023-2024: Emma Karoune (EK), Malvika Sharan (MS) \\
    \texttt{6} & Turing's Partnership in Innovate UK BridgeAI Programme & 2023 & 2024 - Ongoing & Engage and Involve Members \textbf{(Level 3)} - Implementation \textbf{(Stage 3)} & \href{https://www.turing.ac.uk/partnering-turing/current-partnerships-and-collaborations/innovateukbridgeai}{\emph{Details}} & 2024-Present: Alexandra Araujo Alvarez (AAA),	Kirstie Whitaker (KW) \\
    \texttt{7} & \emph{The Turing Way} Practitioners Hub & 2023 & 2023 - Ongoing & Invite Community Feedback \textbf{(Level 2)} - Growth and Scaling \textbf{(Stage 4)} & \href{https://www.turing.ac.uk/turing-way-practitioners-hub}{\emph{About}}, \href{https://zenodo.org/communities/the-turing-way-practitioners}{\emph{Case studies}} & 2023-Present: Malvika Sharan (MS),	Kirstie Whitaker (KW), Arielle Bennett (AB), Previous RPM Alexandra Araujo Alvarez (AAA) (2022-2023) \\
    \texttt{8} & Data-Centric Engineering (DCE) & 2018 & 2023 - Ongoing & Inform Community/ Invite Community Feedback \textbf{(Level 2)} - Planning and Design/ Implementation \textbf{(Stage 3)} & \href{https://www.turing.ac.uk/research/research-programmes/data-centric-engineering}{\emph{Details}} & 2023-Present: Gabin Kayumbi (GK) \\
    \texttt{9} & Environment and Sustainability (E\&S) Grand Challenge & 2023	& 2024 - Ongoing & Inform Community/Invite Community Feedback \textbf{(Level 2)} - Planning and Design \textbf{(Stage 2)} & \href{https://www.turing.ac.uk/research/interest-groups/environment-and-sustainability}{\emph{Details}} & 2024-Present: Cassandra Gould Van Praag (CGVP) \\
    \texttt{10} & People in Data & 2024 & 2024 - Ongoing & Inform Community/ Invite feedback \textbf{(Level 2)} - Initiation \textbf{(Stage 1)} & \href{https://www.turing.ac.uk/research/research-projects/people-data}{\emph{Details}} & 2024-Present: Emma Karoune (EK) \\
\bottomrule
\end{tabular}\\[10pt]
\begin{center}
      \emph{\textbf{Supplementray 2}: This table summarises various projects at the Turing in which RCM team members are involved as shared in Table 1. The last column provides additional details about the manuscript authors who are /were RCM/SRCMs, Senior Researchers and project leaders with a community focus on the project as of August 2024.} 
\end{center}
\end{table*}


\begin{table*}
\small\sf\centering
\caption*{\textbf{Supplementary 3.1}}
\hypertarget{Sup3-1a}{}
\setlength{\tabcolsep}{3pt} 
\renewcommand{\arraystretch}{1.5} 
\begin{tabular}{p{0.1\linewidth} | p{0.17\linewidth} | p{0.17\linewidth} | p{0.17\linewidth} | p{0.17\linewidth} | p{0.17\linewidth}}
\toprule
Project lifecycle stage & 
\multicolumn{2}{|c|}{RCM's Core Competencies - High proficiency} & 
\multicolumn{3}{|c|}{RCM's Peripheral Competencies - Responsibilities shared} \\
\hline
\midrule
    \texttt{\textbf{Skills mapped to project stage}} &	\textbf{Communications}	& \textbf{Engagement}	& \textbf{Strategic contributions}	& \textbf{Technical skills}	& \textbf{Accountability} \\
    \texttt{Initiation} &	Creating community participation guidelines and templates for community use	&	Stakeholder mapping and creating engagement plan	&	Horizon scanning, identifying opportunities and sharing those through the implementation of communication and engagement strategies	&	Supporting the adoption of reproducibility practices such as through the adoption of appropriate license types, version control system and communications platforms (often led by Research Software Engineers/RSEs and Research Data Scientists/RDSs)	&	Following project timeline and budget, identifying risks and creating mitigation plan for a community project \\
    \texttt{Initiation}	& Communication planning and community content development	& Establishing community engagement strategies	& Facilitating the development of community vision and priorities in line with the project strategy	& Establishing and maintaining community infrastructure and identifying skills needs for the community involvement	& Sharing operational plans for project and community \\
    \texttt{Planning}	& Developing technical documentation	& Providing project roadmap for community engagement	& Establishing team and community-wide ways of working	& Planning and supporting the adoption and implementation of open science strategy (open source, open access, open data)	& Contributing to the planning of team and community meetings to make them inclusive and collaborative \\
    \texttt{Planning}	& Open research communication and dissemination plans	& Organising knowledge sharing activities and workshops for training/skill building	& Implementing institutional policy, Code of Conduct and conflict resolution approaches 	& Contributing to the development and implementation of best practices for data access and management practices (often led by data stewards and data wranglers)	& Contributing to task and resource prioritisation in community in line with project's priorities and ensuring Information management for community access \\
    \texttt{Design}	& Community infrastructure such as process documentation, authorship guidelines, and writing templates	& Designing feedback strategy to engage and gather inputs from the community	& Establishing benchmarks to evaluate the success and effectiveness of the project's efforts in engaging with and benefiting the community	& Testing and identifying ideal tools for the use in the project, introducing the use through technical training and upskilling workshops (alongside RSEs and Skills team)	& Coordinating with the community and contributing to the project-wide reporting \\
    \texttt{Design}	& Curation of project resources and sharing them openly with others	& Designing onboarding, mentoring and upskilling opportunities for community members, elevating them to visible roles	& Consulting other communities and establishing collaboration where possible (to avoid doing the same work)	& Supporting the reuse of software / code / infrastructure (alongside RSE/RDS)	& Contributing to the development and implementation of ethics documentation in line with the institutional process (researchers and ethicist) \\
    \texttt{Design}	& Internal communication channel and social media management guidelines	& Advocating for and integrating EDIA in community and organisational policies	& Contributor, user, stakeholder and public advocacy in the project 	& Maintaining software/code/infrastructure for future usage (alongside RSE/RDS)	& Leading community event planning and coordinating logistical requirements with different teams (event teams) \\
\bottomrule
\end{tabular}\\[10pt]
\end{table*}

\begin{table*}
\small\sf\centering
\caption*{\textbf{Supplementary 3.1 Continues ...}}
\hypertarget{Sup3-1b}{}

\setlength{\tabcolsep}{3pt} 
\renewcommand{\arraystretch}{1.5} 
\begin{tabular}{p{0.15\linewidth} | p{0.15\linewidth} | p{0.15\linewidth} | p{0.15\linewidth} | p{0.15\linewidth} | p{0.17\linewidth}}
\toprule
Project lifecycle stage & 
\multicolumn{2}{|c|}{RCM's Core Competencies - High proficiency} & 
\multicolumn{3}{|c|}{RCM's Peripheral Competencies - Responsibilities shared} \\
\hline
\midrule
    \texttt{\textbf{Skills mapped to project stage}} &	\textbf{Communications}	& \textbf{Engagement}	& \textbf{Strategic contributions}	& \textbf{Technical skills}	& \textbf{Accountability} \\
    \texttt{Implementation}	& External and social platform management and monitoring	& Improving visibility and recognition for the community members	& Creating opportunities and improving rewards, incentives and support structure for the community	& Promoting best practices for data analysis and visualisation (alongside RSE/RDS)	& Raising awareness of resources, financial and budget plans that impact the community (project management team) \\
    \texttt{Implementation}	& Implementation of best practices for communication in and about the project	& Creating opportunities for leadership and onboarding diverse community members on those roles	& Identifying growth opportunities and establishing connections between different projects	& Contributing to platform/product management and technology transfer (alongside RAM)	& Planning stakeholder meetings, managing calendars for community access and inviting contributors from the community \\
    \texttt{Implementation}	& Organising community opportunities and events to allow better communications within the community	& Creating multiple engagement paths to collaboration, providing community spaces and opportunities for discussions	& Promoting open leadership practices in the community, and managing interns, staff or teams in senior community roles	& Providing technical Support such as for code Review (alongside RSE/RDS)	Sharing project Organogram (team structure), project governance and project stakeholder engagement opportunities \\
    \texttt{Growth}	& Establishing community-level governance, in reach and outreach process	& Applying open collaboration and participatory approaches	& Co-creating sustainability plans for the community such as by contributing to funding applications, resource identifications and future development opportunities	& Promoting design approaches for improving user experience (alongside UX specialist and RSE)	& Contributing to the recruitment of community members and upholding community participation and involvement policies, as well as following legal requirements and data protection processes (project manager) \\
    \texttt{Monitoring/ optimisation}	& Organisation level impact assessment and reporting	& Establishing community survey, data analysis and reporting (measuring and improving community engagement, health and infrastructure)	& Promoting the project externally and supporting strategic partnerships	& Supporting the development and use of reproducible workflows/research process	& Contributing to internal and external project reporting \\
    \texttt{Sustainability}	& Supporting translation of community resources, sustainability plans, localisation in multiple languages and communications for new users	& Facilitating community involvement in the project	& Supporting the project and community sustainability or sunsetting plans	& Supporting community members in managing, sharing and archiving data and project outputs (alongside data stewards)	& Updating Content Resource Management records for auditing and future references \\
\bottomrule
\end{tabular}\\[10pt]
\begin{center}
    \emph{\textbf{Supplementray 3.1}: RCM Skills and Competency Framework with a detailed version of 65 skills across all competencies.} 
\end{center}
\end{table*}


\begin{table*}
\small\sf\centering
\caption*{\textbf{Supplementary 3.2}}
\hypertarget{Sup3-2}{}

\setlength{\tabcolsep}{3pt} 
\renewcommand{\arraystretch}{1.5} 
\begin{tabular}{p{0.09\linewidth} | p{0.09\linewidth} | p{0.09\linewidth} | p{0.09\linewidth} | p{0.1\linewidth} | p{0.1\linewidth} | p{0.09\linewidth} | p{0.09\linewidth} | p{0.09\linewidth} | p{0.09\linewidth} | p{0.09\linewidth} | p{0.09\linewidth}}
\toprule
\multicolumn{2}{|c|}{Skills Area 1} & 
\multicolumn{2}{|c|}{Skills Area 2} & 
\multicolumn{2}{|c|}{Skills Area 3} & 
\multicolumn{2}{|c|}{Skills Area 4} & 
\multicolumn{2}{|c|}{Skills Area 5} \\
\hline
\midrule
    \textbf{Content} & \textbf{Communi-cation} & \textbf{Engage-ment}	& \textbf{Interperso-nal}	& \textbf{Strategic}	& \textbf{Program development}	& \textbf{Technical}	& \textbf{Technical}	& \textbf{Business}	& \textbf{Program management} \\
    \emph{Community roundtable}	& \emph{CSCCE} &	\emph{Community roundtable}	& \emph{CSCCE} & \emph{Community roundtable}	& \emph{CSCCE} & \emph{Community roundtable}	& \emph{CSCCE} &\emph{Community roundtable}	& \emph{CSCCE} \\
    Comms planning	& Content planning	& Listening \& analysing	& Engagement	& Community strategy	& Strategy development	& Systems admin \& configuration	& Media production	& Program management	& Operational planning \& implementation \\
    Writing	& Content creation \& curation	& Response \& Escalation	& Collaboration	& Roadmap development	& Analysis	& Data collection \& analysis	& Data analysis	& Business Model development	& Time management \\
    Graphics \& Design	& Editorial	& Moderation \& Conflict facilitation	& Training \& teaching	& Policy \& guideline development	& Synthesis	& Tool evaluation \& recommendation	& Data visualisation	& Budget \& Financial Management	& Record-keeping \\
    Multimedia production	& Marketing and branding	& Promoting productive behaviours	& Networking	& Needs \& Competitive analysis	& Proposal development	& Technical support	& Data Management	& Team Hiring \& Management	& Reporting \\
    Narrative development	& Knowledge brokering	& Empathy \& member support	& Coaching \& mentoring	& Measurement, benchmarking \& reporting	& Advancement, growth \& sustainability	& Member Database Management	& Systems administration \& maintenance	& Contractor hiring and management	& Evaluation \& assessment \\
    Editing	& Media relations	& Facilitating connections	& Moderation, mediation \& intervention	& Trendspotting \& Synthesizing	& Advocacy	& Platform architecture \& integration	& Platform product management	& Selling, influencing \& evangelizing	& Event planning \\
    Curation	& Outreach	& New member recruitment	& Emotional integration	& Consulting	& Program design	& Technology issue resolution	& Web and UI design	& Community advocacy \& promotion	& Financial management \\
    Program \& event planning	& Speaking \& Presenting	& New member welcoming	& Cultural competence	& Content strategy development	& Change management	& Software \& application programming	& Content management system admin	& Training development \& delivery	& Community governance \\
    Taxonomy \& Tagging Management	& Social Media	Member & Advocacy	& Consultation \& Listening	& Evaluating engagement techniques	& Recruitment	& UX \& Design	& Technical Support	& Vendor management	& Meeting facilitation \\
    SEO or internal search optimisation	 & -	& Behaviour change \& gamification	 & -	& -	 & -	& Algorithm design \& Data Manipulation	 & -	& Governance management	& - \\
    - &  -	& Training development \& delivery	&  -	&  -	&  -	&  -	&  -	&  -	&  - \\
    -	&  -	& Meeting facilitation	&  -	&  -	&  -	&  -	&  -	&  -	&  - \\
\bottomrule
\end{tabular}\\[10pt]
\begin{center}
      \emph{\textbf{Supplementray 3.2}: Comparative analysis of skills across two reference frameworks used in the article.	Reference 1: \href{https://communityroundtable.com/what-we-do/models-and-frameworks/community-skills-framework/}{The Community Roundtable Community Skills Framework}. Reference 2: CSCCE	Woodley, Lou, Pratt, Katie, Sandström, Malin, Wood-Charlson, Elisha, Davison, Jennifer, \& Leidolf, Andreas. (2021). The CSCCE Skills Wheel – Five core competencies and 45 skills to describe the role of the community engagement manager in STEM. Zenodo. \href{https://doi.org/10.5281/zenodo.4437294}{doi.org/10.5281/zenodo.4437294}}
\end{center}
\end{table*}


\begin{figure*}
\centering
\includegraphics[width=1\textwidth]{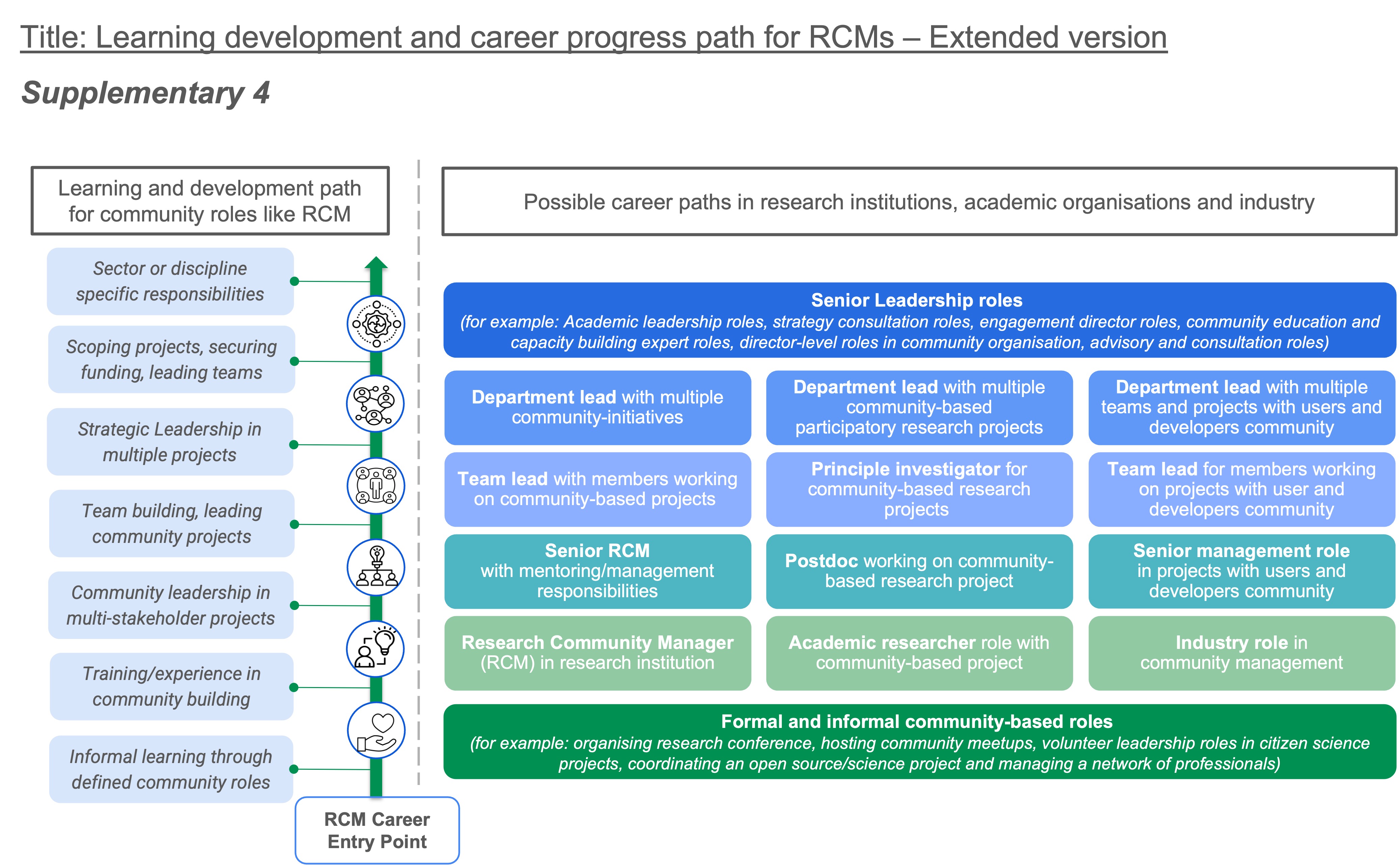}
\caption*{\textbf{Learning development and career progress path for RCMs – Extended version}}
\hypertarget{Sup4}{}
\end{figure*}

\end{document}